\documentclass[
aps,
pra,
superscriptaddress,
showpacs,
showkeys,
notitlepage,
twocolumn, 
numbers,
longbibliography,
nofootinbib]
{revtex4-1}

\usepackage[T1]{fontenc} 
\usepackage[utf8]{inputenc} 
\usepackage[english]{babel} 
\usepackage{graphicx, xcolor}
\usepackage{amsfonts, amssymb, amsmath, mathrsfs, calc, array, mathtools}
\usepackage[separate-uncertainty=true]{siunitx}
\usepackage{hyperref}
\usepackage{abraces}
\usepackage{subcaption}
\usepackage{tikz}
\usetikzlibrary{quantikz}
\usepackage{algorithm,algpseudocode}

\def\bra#1{\mathinner{\langle{#1}|}}
\def\ket#1{\mathinner{|{#1}\rangle}}

\def\ontop#1#2{\setbox0\hbox{#2}\copy0\llap{\raise\ht0\hbox{#1}}}

\newcommand{\defeq}{\vcentcolon=}

\definecolor{darkblue}{rgb}{0,0,0.93} 
\definecolor{darkred}{rgb}{0.8,0,0} 
\definecolor{darkgreen}{rgb}{0,0.7,0}

\hypersetup{
colorlinks=true, 
linkcolor=darkblue, 
citecolor=darkred, 
filecolor=darkblue, 
urlcolor=darkblue
}

\begin{document}

\title{Adjustable-depth quantum circuit for position-dependent coin operators \\ of discrete-time quantum walks}

\author{Ugo Nzongani}
\email{ugo.nzongani@universite-paris-saclay.fr}
\affiliation{Universit{\'e} Paris-Saclay, CNRS, ENS Paris-Saclay, INRIA, Laboratoire M{\'e}thodes Formelles, 91190 Gif-sur-Yvette, France}









\author{Pablo Arnault}
\email{pablo.arnault@inria.fr}
\affiliation{Universit{\'e} Paris-Saclay, CNRS, ENS Paris-Saclay, INRIA, Laboratoire M{\'e}thodes Formelles, 91190 Gif-sur-Yvette, France}




\begin{abstract}

Discrete-time quantum walks with position-dependent coin operators have numerous applications.
For a position dependence that is sufficiently smooth, it has been provided in Ref.\ \cite{NZDPplus2022} an approximate quantum-circuit implementation of the coin operator that is efficient.
If we want the quantum-circuit implementation to be exact (e.g., either, in the case of a smooth position dependence, to have a perfect precision, or in order to treat a non-smooth position dependence), but the depth of the circuit not to scale exponentially, then we can use the linear-depth circuit of Ref.\ \cite{NZDPplus2022}, which achieves a depth that is linear at the cost of introducing an exponential number of ancillas.
In this paper, we provide an adjustable-depth quantum circuit for the exact implementation of the position-dependent coin operator.
This adjustable-depth circuit consists in (i) applying in parallel, with a linear-depth circuit, only certain packs of coin operators (rather than all of them as in the original linear-depth circuit \cite{NZDPplus2022}), each pack contributing linearly to the depth, and in (ii) applying sequentially these packs, which contributes exponentially to the depth.

\end{abstract}

\keywords{}

\pacs{}

\maketitle


\section{Introduction}

Quantum walks are models of quantum transport on graphs \cite{Kempe2003a, Venegas_review}.
They exist both in continuous time \cite{FG98a} and in discrete time \cite{book_FH65, ADZ93a, BB94a}.
In terms of computer science, quantum walks are a model of computation, which has been shown to be universal, both in the continuous- \cite{Childs2009, CGW13} and in the discrete-time case \cite{Lovett2010}, that is, any quantum algorithm can be written in terms of a quantum walk; moreover, many algorithms solving a variety of tasks have been conceived with quantum walks \cite{AKR2005, Amb07a, Tulsi2008, RGAM20, FZAD22}.
In terms of physics, quantum walks are particularly suited to simulate quantum partial differential equations such as the Schrödinger equation or the Dirac equation \cite{Strauch06a, Shikano2013, Arrighi_higher_dim_2014, DAP16, APAF18, DMA19, Arnault2022, Debbasch2019a, AC22} -- the latter being the equation of motion for matter particles which are both quantum and relativistic --, or models of solid-state physics \cite{APP20}.

Discrete-time quantum walks (DQWs), in particular, are discretizations of the Dirac equation which respect both unitarity -- as continuous-time quantum walks (CQWs) -- and strict locality of the transport (contrary to CQWs), that is, concerning the latter point, they preserve relativistic locality on the lattice \cite{AFF14a, BDAP16, Debbasch2019a, Debbasch2019b}.
These DQWs combine, as basic ingredients, shifts on the lattice which depend on the internal state of the particle, together with internal-state rotations, called coin operators, that ``reshuffle the cards'' regarding whether one goes in one direction or another.
The Dirac equation coupled to a variety of gauge fields has been shown to be simulatable with DQWs having a coin operator which depends on the position of the walker on the graph \cite{DMD13b, DMD14, AD16a, AD16b, AMBD16, AD17, AF17, ADMplus2018, MMAMP18, JAD21}.
Position-dependent coin operators also arise when considering randomly chosen coin operators, which are a model of noise in DQWs \cite{Kendon2003, Ken07a, SCPGplus2011,  Ahlbrecht2012, DiMolfetta2016, AMACplus20}.

In Ref.\ \cite{NZDPplus2022}, we have presented different quantum circuits that achieve the implementation of a DQW on the line with such a position-dependent coin operator.
In this paper, we propose a family of quantum circuits with adjustable depth, parametrized by a parameter $m\in\mathbb{N}$, the extremes of which are (i) the naive circuit of Ref.\ \cite{NZDPplus2022} for $m=0$, which  means that all coin operators are implemented sequentially, and (ii) the linear-depth circuit of Ref.\ \cite{NZDPplus2022} for $m=n$ (where $n$ is the number of qubits used to encode the position of the walker in base 2), which means that all coin operators are implemented in parallel.
A higher (lower) $m$ means that more (less) coin operators are implemented in parallel, so that one can choose $m$ as best suited for the experimental platform, knowing that a higher $m$, and hence a smaller depth, requires more ancillary qubits.

In Sec.\ \ref{sec:framework}, we recall the system on which we work, which is that of Ref.\ \cite{NZDPplus2022}, namely, a DQW on the cycle with $N=2^n$ nodes, $n \in \mathbb{N}$.
Such a DQW is made of two operations: a coin rotation $C^{(n)}$, and a coin-dependent shift operation $S^{(n)}$.
Still in Sec.\ \ref{sec:framework}, we recall how to implement $S^{(n)}$ with a quantum circuit.
In Sec.\ \ref{sec:adjustable-depth_circuit}, we introduce our adjustable-depth quantum circuit for the implementation of the position-dependent coin rotation $C^{(n)}$.
The idea of this circuit is to (i) apply in parallel, with a linear-depth circuit such as that introduced in Ref.\ \cite{NZDPplus2022}, only certain packs of coin operators (rather than all of them as in the original linear-depth circuit \cite{NZDPplus2022}), each pack contributing linearly to the depth, and to (ii) apply sequentially these packs, which contributes exponentially to the depth.
In Sec.\ \ref{sec:implementation}, we implement our adjustable-depth quantum circuit on IBM's QASM, the classical simulator of IBM's quantum processors.
In Sec.\ \ref{sec:conclusions}, we conclude and discuss our results.

\section{Framework}
\label{sec:framework}

\subsection{The walk}

The system we consider is the same as that of Ref.\ \cite{NZDPplus2022}, namely, a DQW on a cycle with $N=2^n$ nodes, $n\in\mathbb{N}$.
Let us briefly recall the features of this system.
Each node, labeled as $k=0,\dots,N-1$, is associated to a position quantum state $\ket{k}$.
Let $\mathcal{H}_{\text{pos.}}$ be the $2^n$-dimensional Hilbert space spanned by the position basis $\{\ket{k}\}_{k=0,\dots,N-1}$.
The quantum state of the DQW has an additional, ``internal'' degree of freedom, which is called the coin.
Such a coin belongs to a two-dimensional Hilbert space $\mathcal{H}_0$, the basis of which  is $(\ket{\uparrow}, \ket{\downarrow})\equiv(\ket{0} ,\ket{1}) \equiv ( (1,0)^{\top}, (0,1)^{\top})$, where $\top$ denotes the transposition.
The total Hilbert space to which the quantum state belongs is therefore $\mathcal{H} \defeq \mathcal{H}_{\text{pos.}} \otimes \mathcal{H}_{0}$.
Such a quantum state at time $j\in \mathbb{N}$ decomposes as follows on the basis of $\mathcal{H}$,
\begin{equation}
\label{eq:state}
\ket{\psi_j} = \sum_{k=0}^{N-1} \left( \psi^{\uparrow}_{j,k} \ket{k}\ket{0} +  \psi^{\downarrow}_{j,k} \ket{k}\ket{1} \right) \, ,
\end{equation}
where the complex numbers $\psi^{\uparrow}_{j,k}$ and $\psi^{\downarrow}_{j,k}$ are the coefficients of the decomposition.

The evolution of the quantum state, Eq.\ \eqref{eq:state}, is governed by the following dynamics,
\begin{equation}
\ket{\psi_{j+1}} = W^{(n)} \ket{\psi_j} \, ,
\end{equation}
where the \emph{walk operator} $W^{(n)}$ is composed of two operations,
\begin{equation}
W^{(n)} \defeq S^{(n)} C^{(n)} \, .
\end{equation}
The first operation is a possibly position-dependent total \emph{coin operator},
\begin{equation}
\label{eq:cn}
C^{(n)} \defeq \sum_{k=0}^{N-1} \ket{k} \! \! \bra{k} \otimes {C}_{k} \, ,
\end{equation}
where each $C_k$ is a coin operator, that is, here, a $2\times 2$ complex matrix acting on $\mathcal{H}_0$.
The second and last operation is a \emph{coin-dependent shift operator},
\begin{equation}
\label{eq:sn}
\begin{split}
S^{(n)} &\defeq \sum_{k=0}^{N-1} \Big( \ket{k-1 \ \text{mod} \ N} \! \! \bra{k} \otimes \ket{0} \! \! \bra{0}   \\
& \ \ \ \ \ \ \ \ \   + \ket{k+1 \ \text{mod} \ N} \! \! \bra{k} \otimes \ket{1} \! \! \bra{1} \Big) \, .
\end{split}
\end{equation}

\subsection{The encoding in base 2}

What is the minimum number of entangled qubits, i.e., of wires, that we need in our quantum circuit in order to encode the $N=2^n$ nodes of the cycle: this minimum number is $n$, and we call these qubits \emph{position qubits}. 
This naturally provides a base-2 encoding of the position of the walker.
Let $\ket{k_2}$ be the writing of $\ket k$ in base 2, that is:
\begin{equation}
\ket k \equiv \ket{k_2} \equiv \ket{b_{n-1} ... b_0} \, ,
\end{equation}
where $b_p = 0$ or $1$ with $p=0, ...,n-1$, such that $k = \sum_{p=0}^{n-1} b_p \times 2^p$.
One can thus rewrite Eqs.\ \eqref{eq:cn} and \eqref{eq:sn} as
\begin{equation}
\label{eq:non-uniform_coin_operator}
{C}^{(n)} \equiv \sum_{k=0}^{2^n-1} \ket{k_2} \! \! \bra{k_2} \otimes \tilde{C}_{k_2}  \, ,
\end{equation}
where
\begin{equation}
\tilde{C}_{k_2} \defeq C_k \, ,
\end{equation}
and,
\begin{equation}
\begin{split}
{S}^{(n)} &\equiv \sum_{k=0}^{2^n-1} \Big( \ket{(k-1 \ \text{mod} \ N)_2} \! \! \bra{k_2} \otimes \ket{0} \! \! \bra{0}   \\
& \ \ \ \ \ \ \ \ \   + \ket{(k+1 \ \text{mod} \ N)_2} \! \! \bra{k_2} \otimes \ket{1} \! \! \bra{1} \Big) \, .
\end{split}
\end{equation}

\subsection{Final remarks}

In order to be able to implement suck a walk $W^{(n)}$ with a quantum circuit, one has to be able to implement the two operators $C^{(n)}$ and $S^{(n)}$.
There are different ways of implementing the coin-dependent shift operator $S^{(n)}$ using a quantum circuit, as recalled in Ref.\ \cite{NZDPplus2022}.
The aim of this paper is the quantum-circuit implementation of a position-dependent total coin operator $C^{(n)}$ with a circuit having a depth that can be adjusted at will, which is the subject of the next section.

\section{Adjustable-depth circuit}
\label{sec:adjustable-depth_circuit}

\subsection{General idea}

\subsubsection{The idea}

As mentioned in the introduction, the general idea of the adjustable-depth circuit we introduce in this paper is to (i) apply in parallel, with a linear-depth circuit such as that introduced in Ref.\ \cite{NZDPplus2022}, only certain packs of coin operators $\tilde{C}_{k_2}$ (rather than all of them as in the original linear-depth circuit \cite{NZDPplus2022}), each pack contributing linearly to the depth, and to (ii) apply sequentially these packs, which contributes exponentially to the depth.

The total number of coin operators $\tilde{C}_{k_2}$ is $2^n$.
The size of the packs, i.e., the number of coin operators that we apply in parallel, is the tunable parameter of our model, and we write it as a power of $2$ to simplify the discussion, that is, we write it $M = 2^m$, $m\in \mathbb{N}$.
The number of packs is thus $2^n/2^m=2^{n-m}$.
Let us call $U_i^{(n,m)}$, $i = 0, ...,2^{n-m}-1$, the circuit that implements the $i$th pack of coin operators $\tilde{C}_{k_2}$ in parallel.
The total circuit, which we are going to show implements the coin operator $C^{(n)}$, thus reads
\begin{equation}
\label{eq:packs}
U^{(n,m)} \defeq  \sideset{}{^L}\prod_{i=0}^{2^{n-m}-1} U_i^{(n,m)} \, ,
\end{equation}
where the superscript $L$ means that the terms are multiplied in increasing index order from right to left. 
The number $i$ is called the stage number.

\subsubsection{The ancillae, and more precisions}

\begin{figure}[t]
\hspace{-0.0cm}
	\includegraphics[width=7cm]{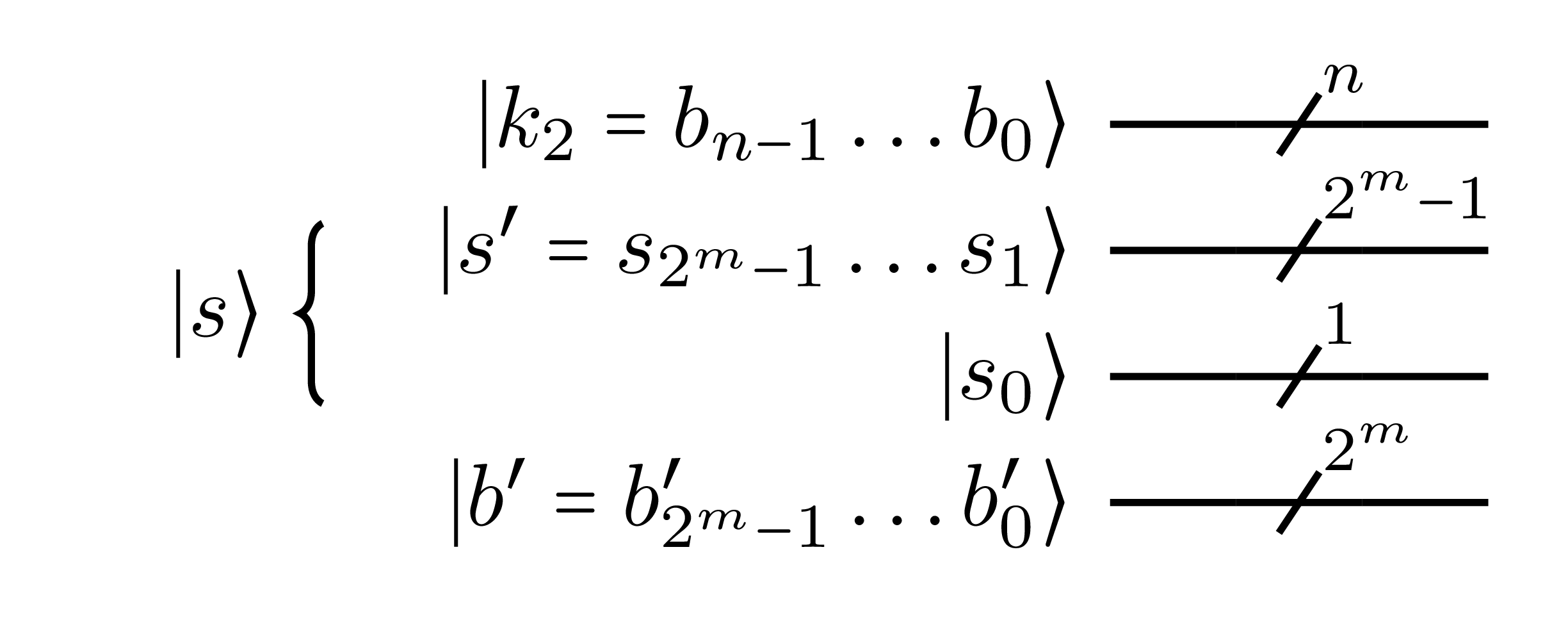}
	\caption{Registers necessary for the implementation of $U^{(n,m)}$. \label{fig:registers}}
\end{figure}

\begin{figure}[t]
\hspace{-0.0cm}
	\includegraphics[width=8.8cm]{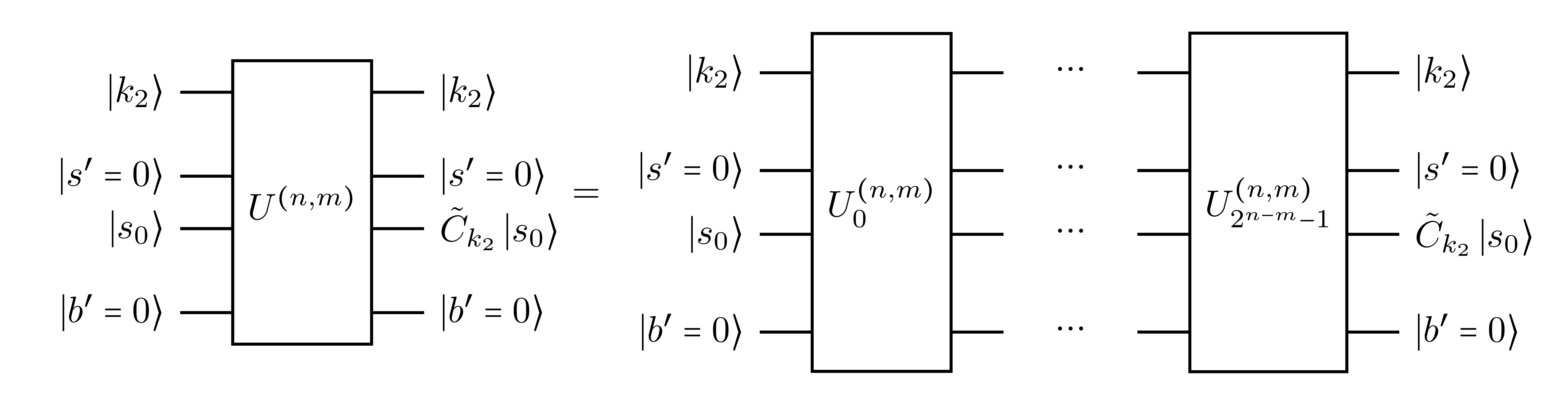}
	\caption{Decomposition of $U^{(n,m)}$ in $2^{n-m}$ packs $U_i^{(n,m)}$, as written in Eq.\ \eqref{eq:packs}. \label{fig:packs}}
\end{figure}

The number of ancillae necessary to apply each $U_i^{(n,m)}$ is: $2^m$ ancillary position states, and $2^m-1$ ancillary coin states.
 We depict in Fig.\ \ref{fig:registers} the different registers used to implement our circuit $U^{(n,m)}$.
In Fig.\ \ref{fig:packs}, we illustrate Eq.\ \eqref{eq:packs}.

As in Eq.\ (26) of Ref.\ \cite{NZDPplus2022}, the fact that $U^{(n,m)}$ does the job, i.e., implements $C^{(n)}$, means that it coincides with $C^{(n)}$ on the Hilbert space spanned by the position qubits plus the coin qubit, provided that we have correctly initialized the ancillary qubits.
We are going to detail this in the next paragraph.

Including the ancillae means extending the total Hilbert space $\mathcal{H}$ introduced in Sec.\ \ref{sec:framework} into $\mathcal{H}' \defeq \mathcal{H}\otimes \mathcal{H}'_{\text{coins}} \otimes \mathcal{H}'_{\text{pos.}}$.
The last two Hilbert spaces contain respectively  the quantum states of the coin and position ancillary qubits.
A correctly initialized quantum state $\ket{S} \in\mathcal{H}'$ is a state that is arbitrary on $\mathcal{H}$, but has to be equal to $\ket{s'=0}\ket{b'=0}$ on $\mathcal{H}'_{\text{coins}} \otimes \mathcal{H}'_{\text{pos.}}$, that is,
\begin{equation}
\label{eq:special}
    \ket S \defeq \left( \sum_{k=0}^{2^n-1} \sum_{s_0=0,1} \alpha_{k,s_0} \ket{k_2}\ket{s_0} \right)  \ket{s'=0} \ket{b'=0} \, ,
\end{equation}
with the $\alpha_{k,s_0}$'s being complex numbers such that $\sum_{k=0}^{2^n-1} \sum_{s_0=0,1} |\alpha_{k,s_0}|^2 = 1$.
As we said above, that $U^{(n,m)}$ does the job, i.e., implements $C^{(n)}$, means the following,
\begin{equation}
\label{eq:the_thing}
U^{(n,m)} \ket{S} = \left( {C}^{(n)} \otimes I_{2^{(2^m-1)}} \otimes I_{2^{(2^m)}}  \right) \ket{S} \, .
\end{equation}

Notice that in Eq.\ \eqref{eq:special} we have chosen to represent, in that order: the state of the position qubits $\ket{k_2}$, the principal coin $\ket{s_0}$, the ancillary coins $\ket{s'}$, and finally the ancillary position $\ket{b'}$.
This choice has been made for a clearer formulation of the equations.
In contrast, in the diagrammatic representations of our circuits, see Figs.\ \ref{fig:registers} and \ref{fig:packs}, the principal coin was placed under the ancillary coins.
This choice has been made for a better visual understanding of the functioning of the circuits.

\subsection{General structure of  $U^{(n,m)}_i$: that of $U^{(n)}_{\mathrm{lin.}}$ of Ref.\ \cite{NZDPplus2022}}\label{subsec2:adjustable-depth_circuit}

\begin{figure}[t]
\hspace{-0.7cm} \
	\includegraphics[width=9.2cm]{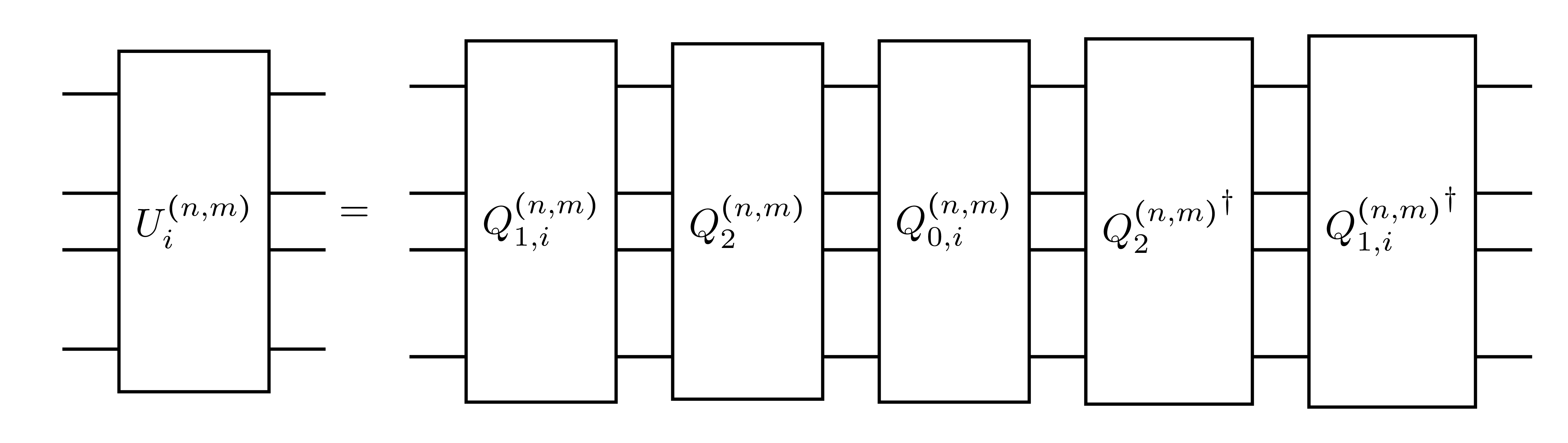}
	\caption{Decomposition of each $U_i^{(n,m)}$, written in Eq.\ \eqref{eq:packsQ}. \label{fig:packQ_1}}
\end{figure}

As the circuit $U^{(n)}_{\text{lin.}}$ in Ref.\ \cite{NZDPplus2022}, each $U^{(n,m)}_i$ is made of several operations; more precisely, it reads
\begin{equation}
\label{eq:packsQ}
    U_i^{(n,m)} \defeq {Q_{1,i}^{(n,m)}}^\dagger {Q_{2}^{(n,m)}}^\dagger Q_{0,i}^{(n,m)}Q_{2}^{(n,m)}Q_{1,i}^{(n,m)} \, .
\end{equation}
Let us first briefly recall the operating principle of $U^{(n)}_{\text{lin.}}$, which is also that of each $U^{(n,m)}_i$: we first encode the ancillary position with $Q_{1,i}^{(n,m)}$; we then swap the state of the principal coin $\ket{s_0}$ onto the ancillary coins via  $Q_{2}^{(n,m)}$; we then apply the running, pack $i$ of coin operators in parallel, via $Q_{0,i}^{(n,m)}$; finally, the ancillary coin states are reset via ${Q_{2}^{(n,m)}}^\dagger$, and the ancillary position states are reset via ${Q_{1,i}^{(n,m)}}^\dagger$.
Equation \eqref{eq:packsQ} is illustrated in Fig.\ \ref{fig:packQ_1}.

\begin{figure*}
\includegraphics[width=4cm]{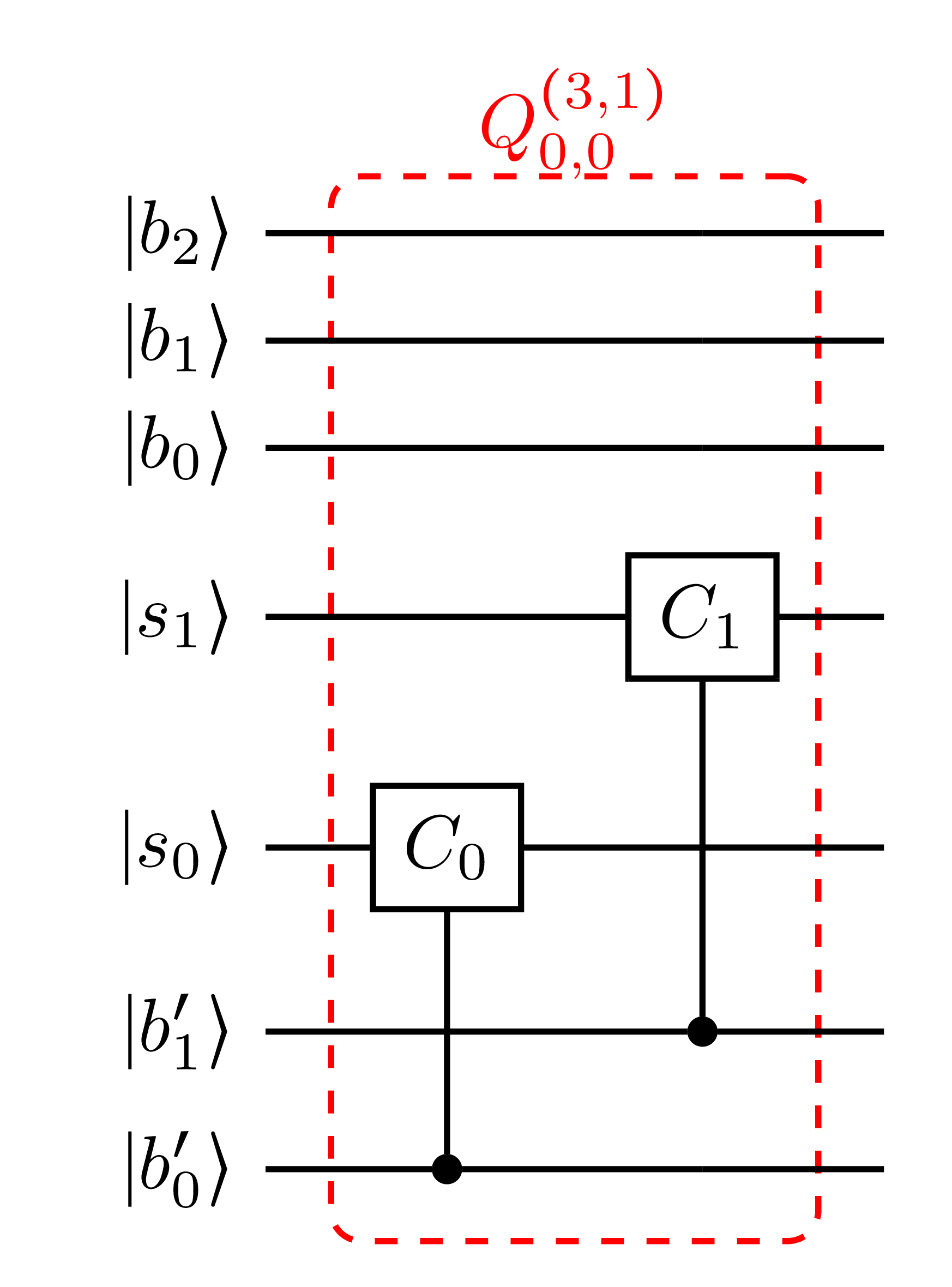}
\includegraphics[width=4cm]{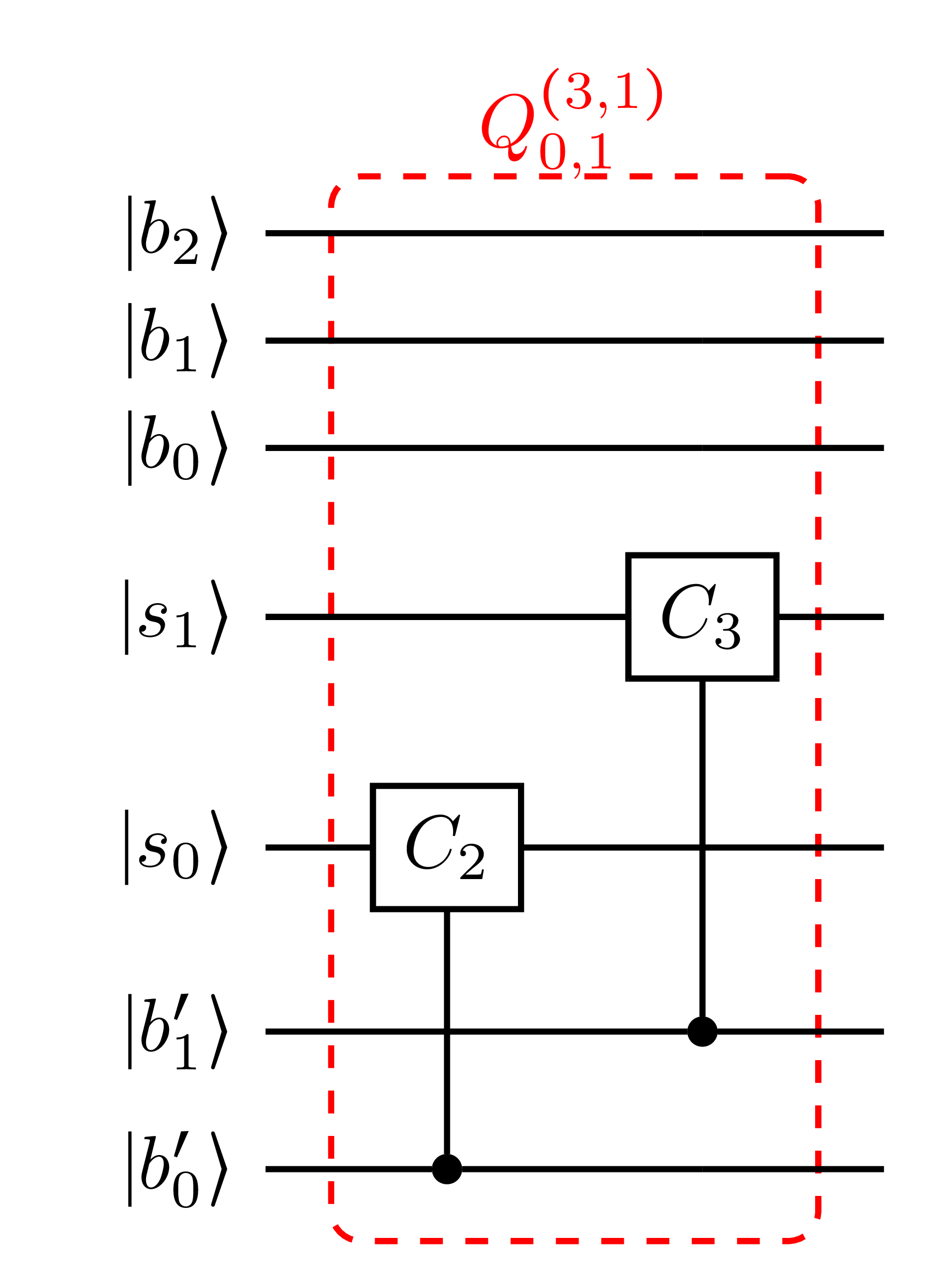}
\includegraphics[width=4cm]{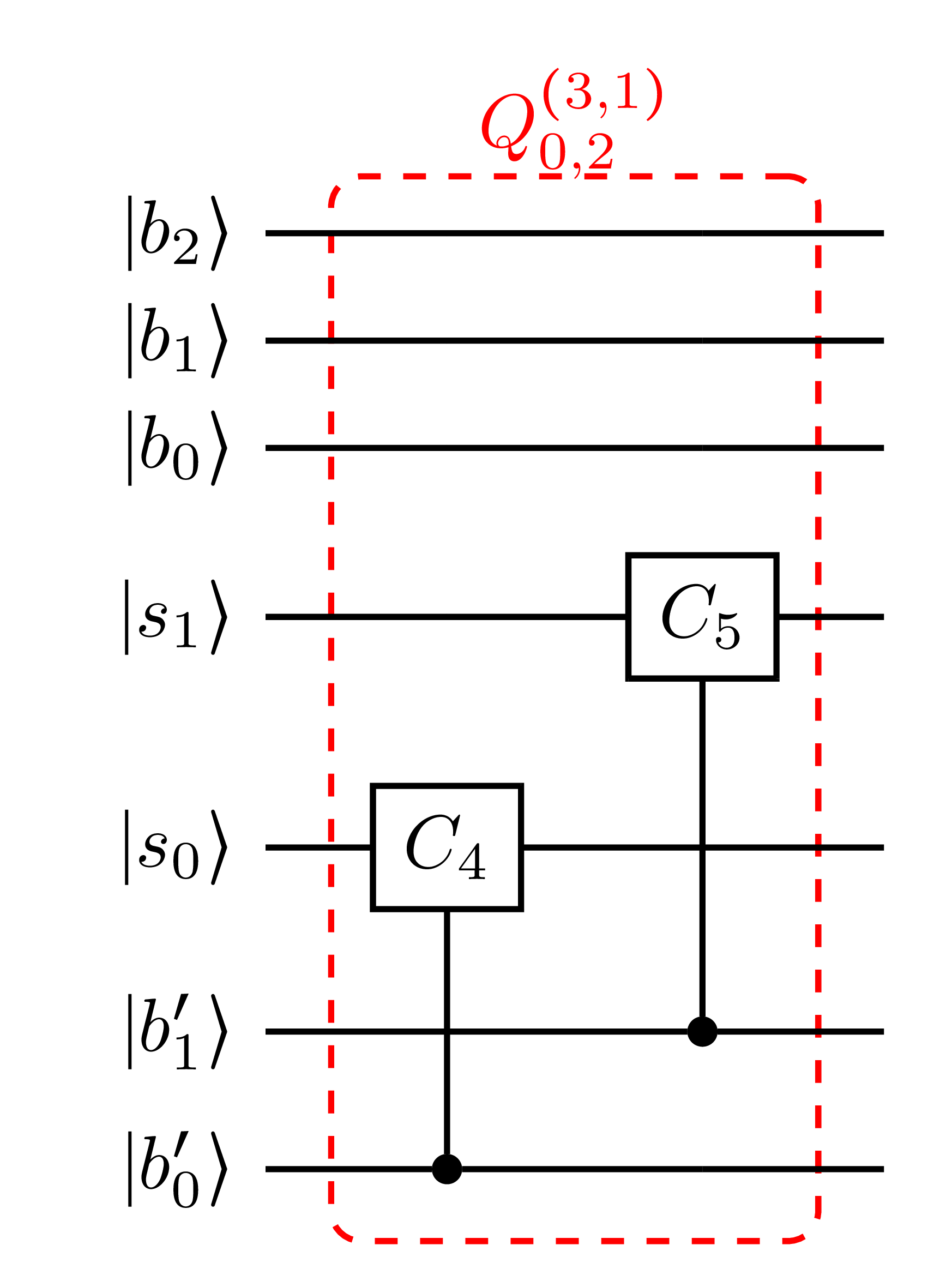}
\includegraphics[width=4cm]{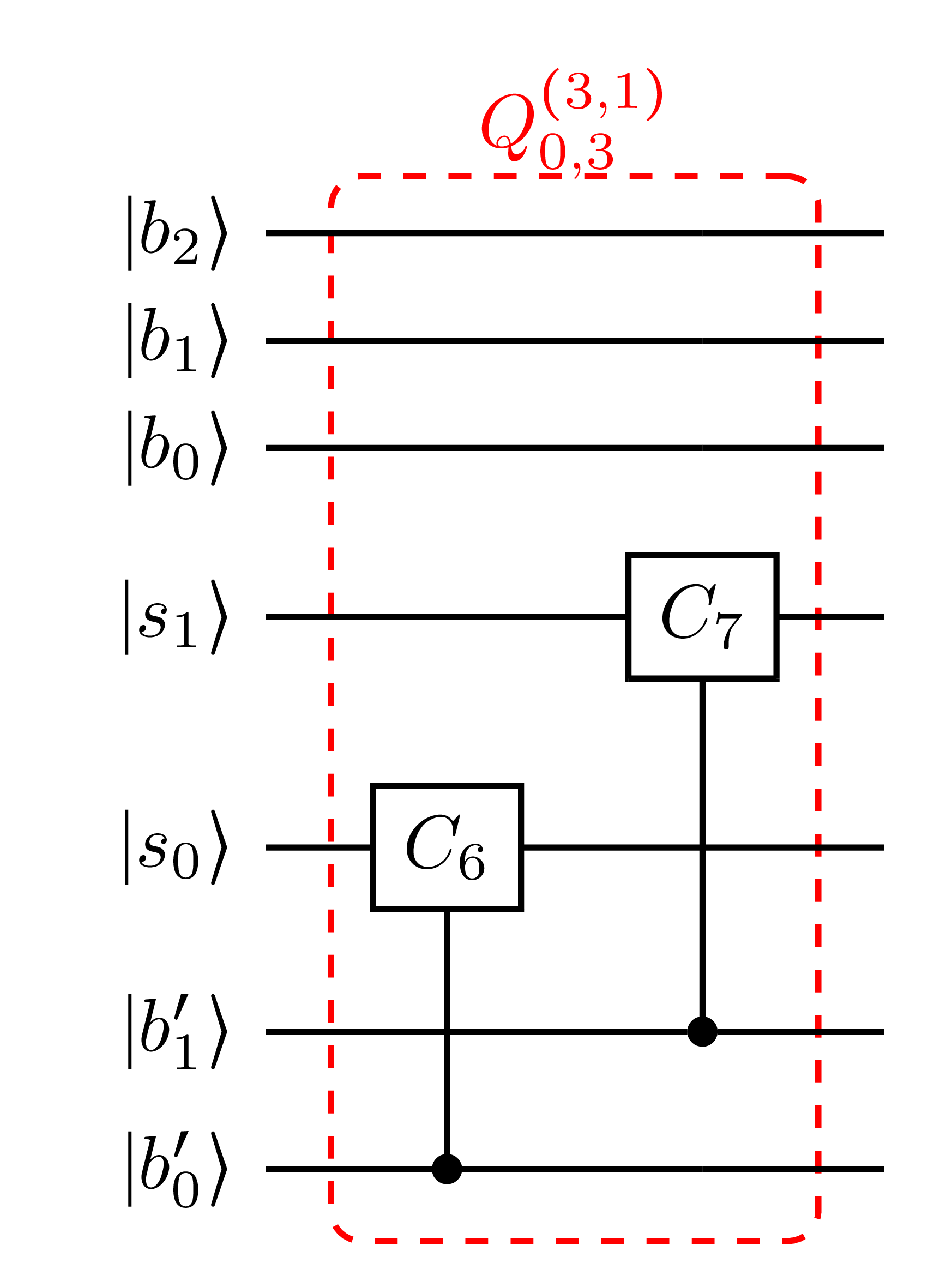}
\caption{Quantum circuits implementing $Q_{0,i}^{(n=3,m=1)}$ for $i=0,1,2,3$, that is, we implement, at each stage, $2^m=2$ coin operators (out of $2^n=8$) in parallel. \label{fig:Q0i}}
\end{figure*}

Now, the central operation, $Q_{0,i}^{(n,m)}$, is the same as $Q_0^{(n)}$ in $U^{(n)}_{\text{lin.}}$ of Ref.\ \cite{NZDPplus2022}, except that we only apply $2^m$ coin operators $\tilde{C}_{k_2}$ in parallel instead of $2^n$.
More precisely, $Q_{0,i}^{(n,m)}$  reads
\begin{equation}
\label{eq:q0}
    Q_{0,i}^{(n,m)} \defeq I_{2^n} \otimes \left(\sideset{}{^L}\bigotimes_{k=0}^{2^m-1} K_{b'_k,s_k}(C_{i2^m+k})\right) \, ,
\end{equation}
where $I_{2^n}$ is applied on the position qubits, and where $K_{a,b}(C)$ corresponds to applying the one-qubit gate $C$ on qubit $\ket b$ while controlling it on qubit $\ket a$ (we apply $C$ only if $a=1$).
In Fig.\ \ref{fig:Q0i}, we illustrate the definition of $Q_{0,i}^{(n,m)}$ in Eq.\ \eqref{eq:q0}.
For $m=n$, we have a single operator $U_0^{n,m=n} = U_{\text{lin.}}^{(n)}$.

\subsection{New ingredient: only in the operator $Q_{1,i}^{(n,m)}$}
\label{subsec:new_ingredient}

\subsubsection{Introduction}

Apart from the fact that we have less ancillae for $m < n$ than for $m=n$, and that $Q_{0,i}^{(n,m)}$ only applies $2^m$ coin operators $\tilde{C}_{k_2}$ in parallel,  the only difference between $U^{(n,m)}_i$ and the circuit $U^{(n)}_{\text{lin.}}$ of Ref.\ \cite{NZDPplus2022}, is in the operation that initializes the ancillary positions, namely, $Q_{1,i}^{(n,m)}$.
Let us explain this difference.

Previously, the operator $Q_1^{(n)}$ of Ref.\ \cite{NZDPplus2022} encoded the ancillary position $\ket{b'}$ for any position $\ket{k_2}$.
Now, as we apply in parallel packs of $2^m$ coin operators only, one must only encode the ancillary position for $2^m$ position states only, not all of them.

We call \emph{current position state at stage $i$} a position state $\ket{k^{(i)}_2}$ such that after stage $i$ the coin operator $\tilde{C}_{k^{(i)}_2}$ has been applied to the principal coin state $\ket{s_0}$.
There are thus $2^m$ current position states at each stage $i$, namely, those for which $k^{(i)}= i2^m, ...,(i+1)2^m-1$ (see Eq.\ \eqref{eq:q0} and Fig.\ \ref{fig:Q0i}).

In Ref.\ \cite{NZDPplus2022}, the ancillary position is encoded for every position state thanks to the fact that the qubit which encodes the least significant bit of the ancillary position, namely $\ket{b_0'}$, is flipped using a NOT gate, before the application of the series of controlled-SWAP operations (see the operation $Q_{11}$ in Fig.\ 15 of Ref.\ \cite{NZDPplus2022}).

\subsubsection{Main explanations}

To encode only the current position states at each stage $i$, one has to flip the same qubit $\ket{b'_0}$ but only for these position states.
Let us explain how to do that.
Let $\ket{k_2} = \ket{b_{n-1} ... b_0}$ be the input position state.
The current position states at any stage $i$ have their last $n-m$ bits in common in their binary writing, which is of the form $k^{(i)}_2 \defeq b_{n-1}^{(i)} ... b_{0}^{(i)}$.
More specifically, it turns out that these $n-m$ bits in common actually code for the binary writing $i_2$ of $i$, that is, we have
\begin{subequations}
\begin{align}
i_2 &\defeq h_{n-m-1} ... h_0 \\
&\defeq b_{n-1}^{(i)} ... b_m^{(i)} \, .
\end{align}
\end{subequations}
Hence, flipping $\ket{b'_0}$ only for the current position states at stage $i$ can be done by controlling the NOT gate with positive (i.e., on $1$) and/or negative (i.e., on $0$) controls on the first $n-m$ position qubits from top to bottom, starting from $\ket{b_{n-1}}$, such that the NOT gate is activated if and only if 
\begin{equation}
b_{n-1} ... b_m = i_2 \, .
\end{equation}
This corresponds to applying a certain generalized $(n-m)$-Toffoli gate, where ``generalized'' means with positive and/or negative controls.
As a reminder, an $n$-Toffoli gate is a NOT gate controlled positively by $n$ qubits. 
Note that a $0$-Toffoli gate thus denotes a NOT gate.
The encoding performed by $Q_{1,i}^{(n,m)}$ is then $\ket{b'=0} \rightarrow \ket{b'=(2^{k^{(i)}-i2^m})_2}$ for the current position states $k^{(i)}$, and $\ket{b'=0} \rightarrow \ket{b'=0}$ (i.e., the identity) for the other position states.

\begin{figure*}[t!]
\includegraphics[width=3.2cm]{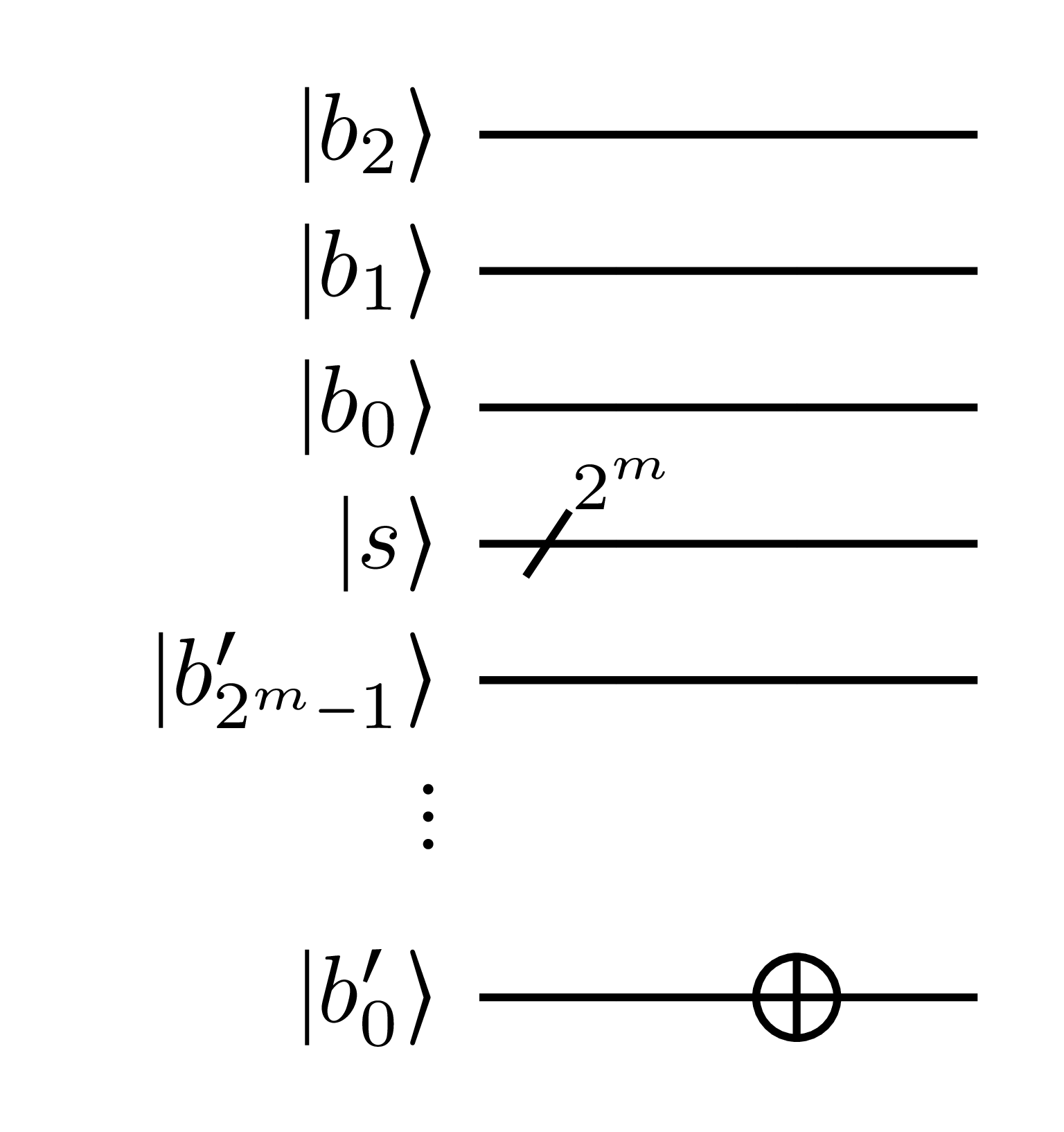} \ \ \
\includegraphics[width=3cm]{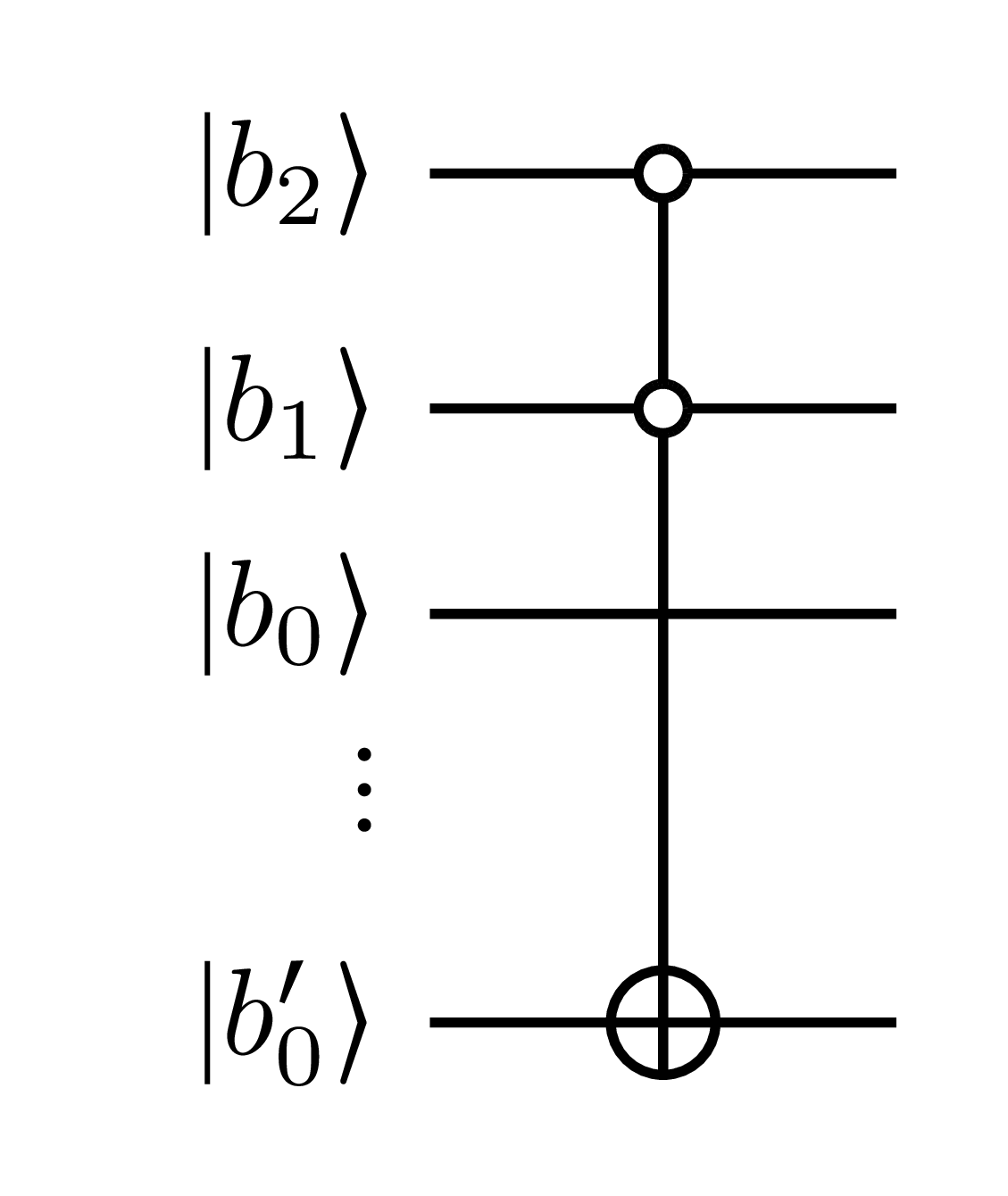} \ \ \
\includegraphics[width=3cm]{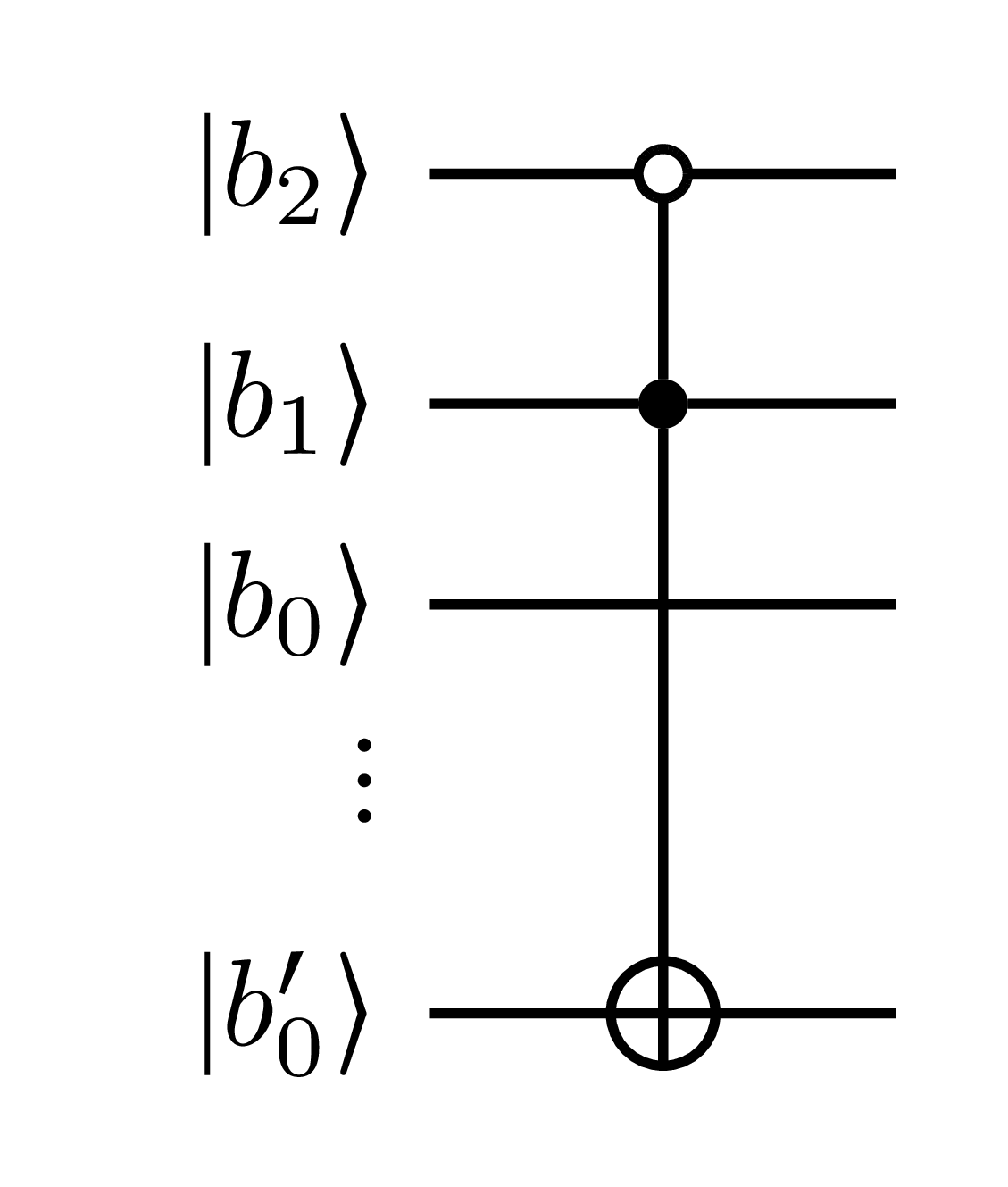} \ \ \
\includegraphics[width=3cm]{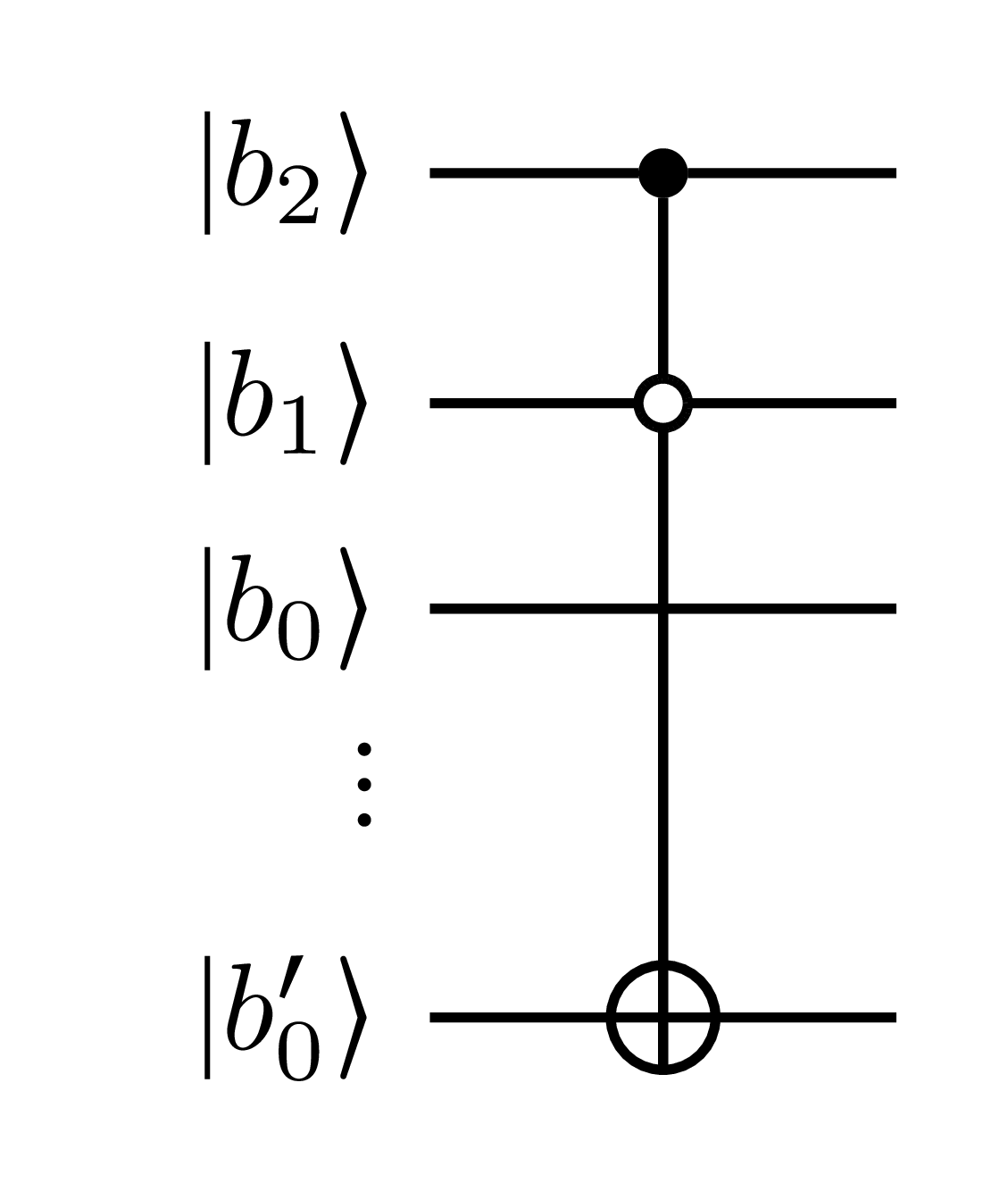} \ \ \
\includegraphics[width=3cm]{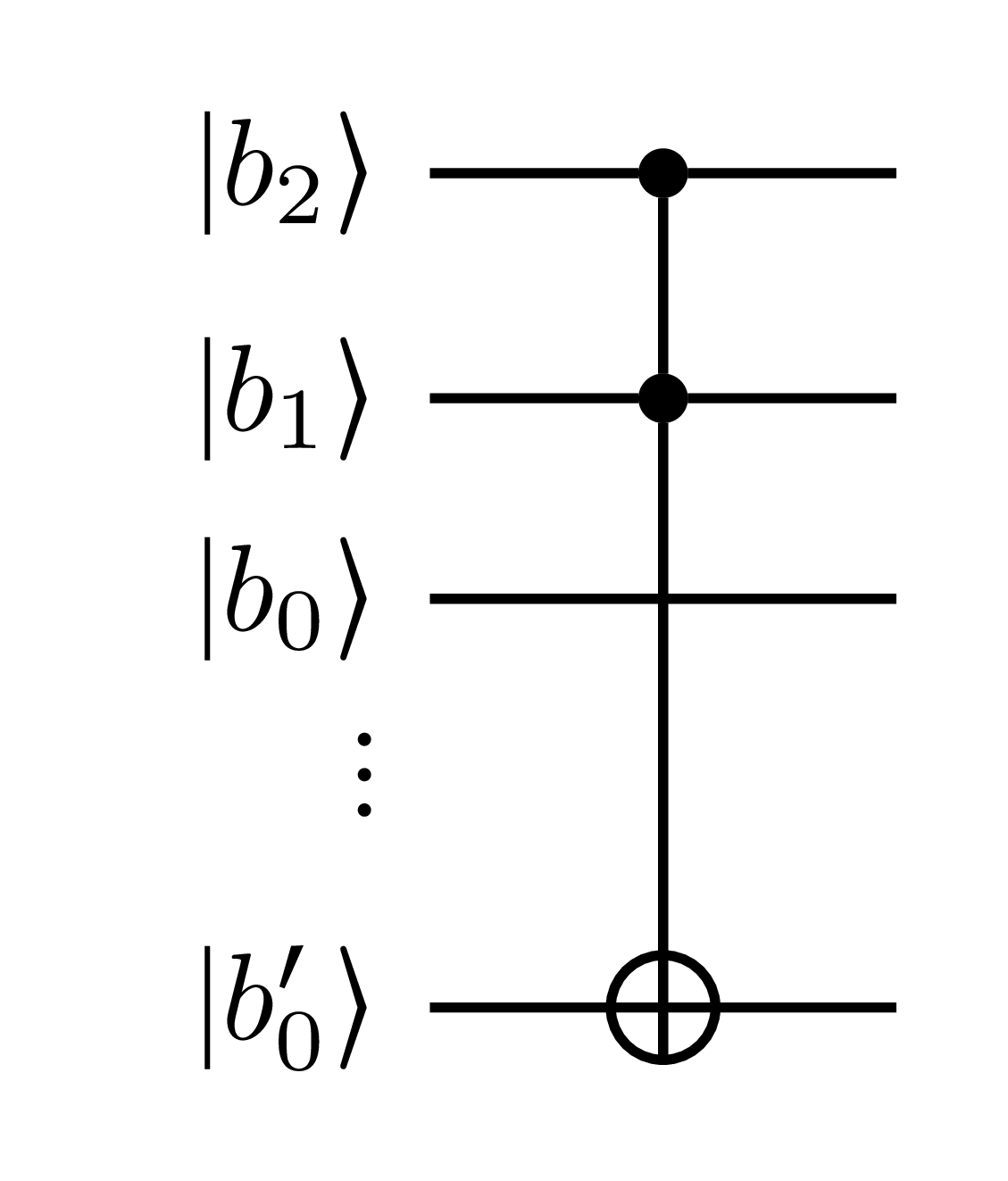}
\caption{We choose $n=3$ and $m=1$. On the far left, there is a circuit with only a NOT gate on $\ket{b'_0}$, which is what would be used at the beginning of the controlled-SWAP operations of each $Q_{1,i}^{(n,m)}$, if there was no difference with $Q_1^{(n)}$ of $U_{\text{lin.}}^{(n)}$ in Ref.\ \cite{NZDPplus2022}. But, here is precisely the main new ingredient of the present adjustable-depth circuit to implement the position-dependent coin operators: instead of this NOT gate, we have to apply, at the beginning of the controlled-SWAP operations of $Q_{1,i}^{(n,m)}$, a generalized $(n-m)$-Toffoli gate which activates the NOT gate if and only if $b_{n-1}...b_m = i_2$, where $i_2$ is the binary writing of $i$. From left to right starting from the second circuit, we have depicted this generalized $(n-m)$-Toffoli gate for $i=0,1,2,3$. The term ``generalized'' simply refers to the fact that the controls can be positive (black dot) or negative (white dot). \label{fig:Toffolis}}
\end{figure*}

In Fig.\ \ref{fig:Toffolis}, we illustrate the above-mentioned generalized $(n-m)$-Toffoli gates of each $Q_{1,i}^{(n,m)}$ for $n=3$ and $m=1$.
These generalized $(n-m)$-Toffoli gates replace the NOT gate at the beginning of $Q_{11}$ in $Q_1$ of $U_{\text{lin.}}^{(n)}$ in Ref.\ \cite{NZDPplus2022}.

In Appendix \ref{app:Q1}, we give the explicit definition of $Q_{1,i}^{(n,m)}$. 
In Appendix \ref{app:Q2}, we, for the sake of completeness, write explicitly $Q_{2}^{(n,m)}$, which initializes the ancillary coins, but as mentioned the only change with respect to $Q_{2}^{(n)}$ of Ref.\ \cite{NZDPplus2022} is the number of ancillary qubits.

In appendices  \ref{app:Q1} and \ref{app:Q2}, we have omitted certain identity tensor factors in certain equations, in order to lighten the writing.

In total, since at each stage the ancillary positions are encoded only if we have a current position state at stage $i$, then it means that the coin operator at a given position is applied only if that position is a current position at stage $i$: in other words, $U^{(n,m)}_i$ coincides on $\mathcal{H}$ with (i) the coin operator at a given position if that given position is a current position at stage $i$, and with (ii) the identity otherwise, which achieves our goal.

\subsection{Depth}

The purpose of this adjustable-depth circuit is that, depending on the parameter $m$, its width and depth will be modified, one for the benefit of the other. Let $w(\cdot)$ and $d(\cdot)$  be respectively the functions that return the width and the depth before compilation of an operator. One also needs to define functions $\varepsilon_w(x)$ and $\varepsilon_d(x)$ which return respectively the number of ancillary qubits that may be needed to implement an $x$-Toffoli gate, and the depth after compilation of the latter. Finally, $\delta_{i,j}$ is the Kronecker symbol.

Since one uses $n$ position qubits, $2^m$ coin qubits and $2^m$ ancillary position qubits, the width of $U^{(n,m)}$ reads
\begin{equation}
\label{eq:width}
    w(U^{(n,m)}) = n + 2^{m+1} + \varepsilon_w(n-m) \, .
\end{equation}

As for the total depth of the circuit, we show in Appendix \ref{app:depth} that it is given by
\begin{equation}
\label{eq:depth}
    d(U^{(n,m)}) = 2^{n-m}(20m+2\varepsilon_d(n-m) + 8\delta_{m,0}-5)-2 \, .
\end{equation}
As shown in Ref.\ \cite{Saeedi_2013}, one can implement an $n$-Toffoli gate linearly in $n$ without using any ancilla.
Therefore, one can consider the term $\varepsilon_d(n-m)$ to be linear in $n-m$, and $\varepsilon_w(n-m)=0$.
We finally remark the exponential dependence in the number of packs $n-m$, namely, $2^{n-m}$, and the linear dependence in the number $m$ of position qubits involved per pack, namely, $20m$ (plus the linear dependence of $\varepsilon_d(n-m) $ in $n-m$).

In Fig.\ \ref{fig:table}, we present the different width and depth complexities of $U^{(n,m)}$ for some remarkable values of $m$, with $N=2^n$.
\begin{figure}
\begin{tabular}{|c|c|c|}
  \hline
  $m$ & \text{Depth complexity} & \text{Width complexity} \\
  \hline
    $m=0$ & $O(\log_2(N)N)$ & $O(\log_2(N))$ \\
  \hline
    $m=\frac{n}{2}$ & $O(\log_2(N) \sqrt{N})$ & $O(\sqrt{N})$ \\
  \hline
    $m=n$ & $O(\log_2(N))$ & $O(N)$  \\
  \hline
\end{tabular}
\caption{Depth and width complexity of our adjustable-depth quantum circuit $U^{(n,m)}$, implementing the position-dependent coin operator, for remarkable values of $m$, where we recall that $N=2^n$. \label{fig:table}}
\end{figure}

\section{Implementation}
\label{sec:implementation}

We have implemented our adjustable-depth quantum circuit on IBM's QASM, the classical simulator of IBM's quantum processors, thanks to the software Qiskit.

\begin{figure}
\includegraphics[width=8.5cm]{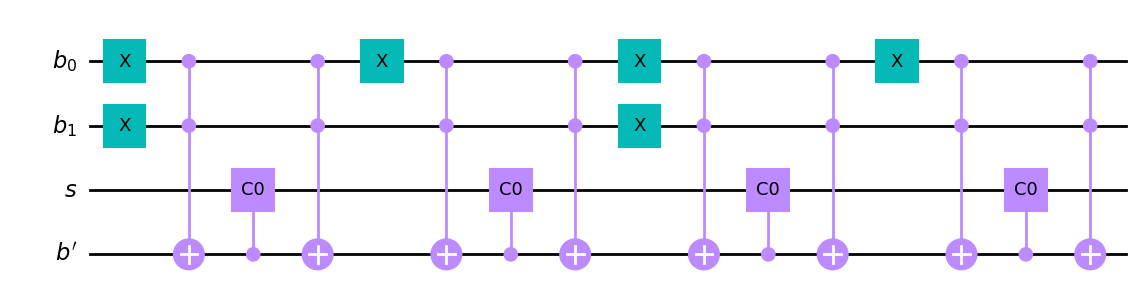} \\
\includegraphics[width=8.5cm]{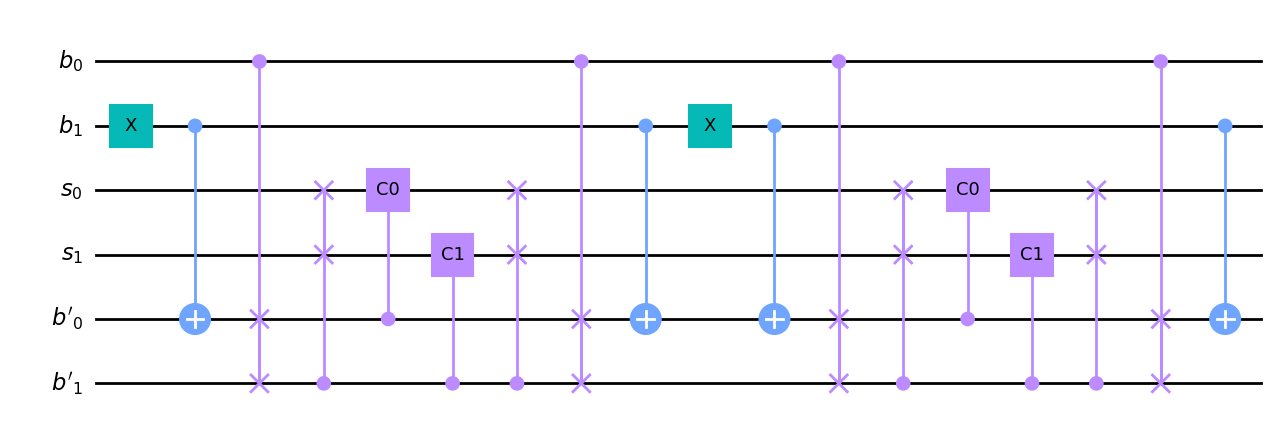} \\\includegraphics[width=8.5cm]{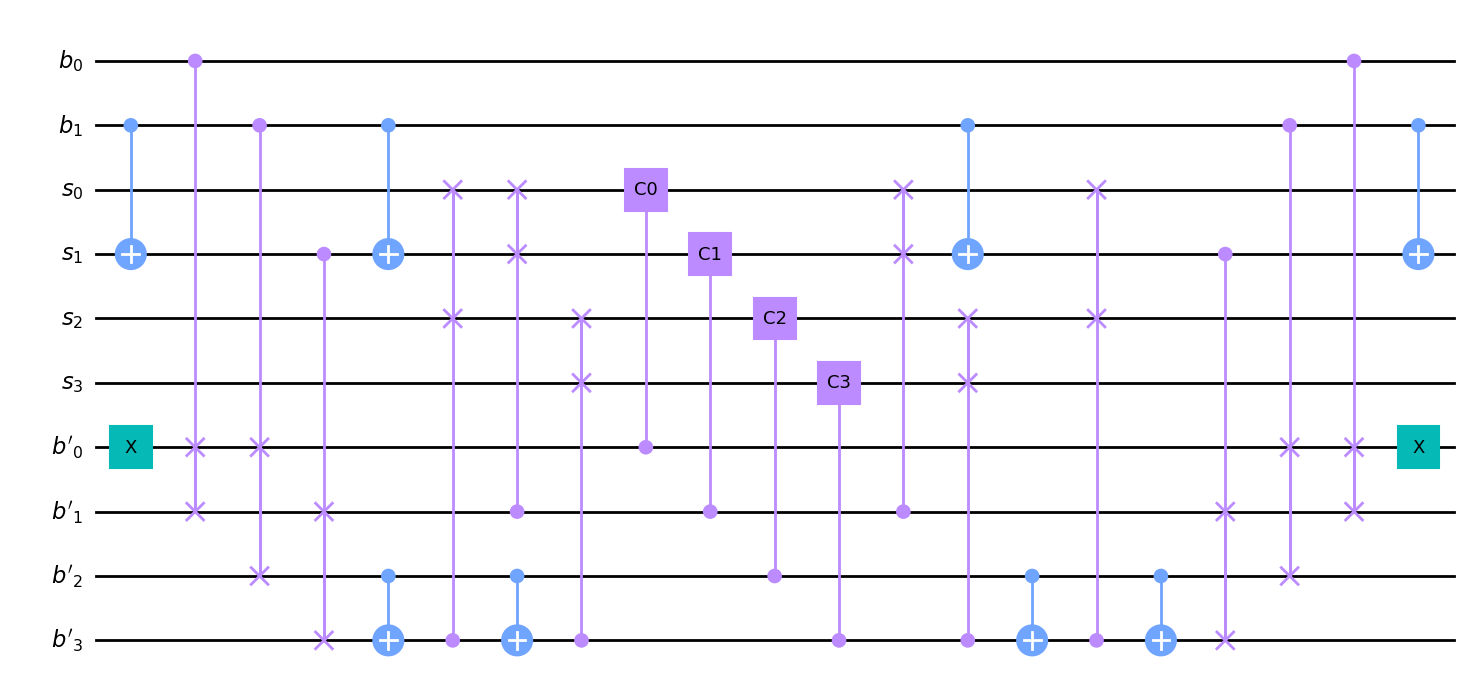}
\caption{\label{fig:qiskit_circuits} Our adjustable-depth quantum circuit for $n=2$ position qubits, and packs of size $2^m$, $m=0$ (top circuit), $1$ (middle circuit), and $2$ (bottom circuit), which means that $2^m$ coin operators are executed in parallel at each of the $2^{n-m}$ stages in the quantum circuit.}
\end{figure}

In Fig.\ \ref{fig:qiskit_circuits}, we show how this quantum circuit looks like for $n=2$ position qubits and an adjustable parameter $m=0,1,2$.
In Appendix \ref{app:pseudo_code}, we give the pseudo-code used in order to generate these circuits.

\begin{figure*}
\includegraphics[width=8.5cm]{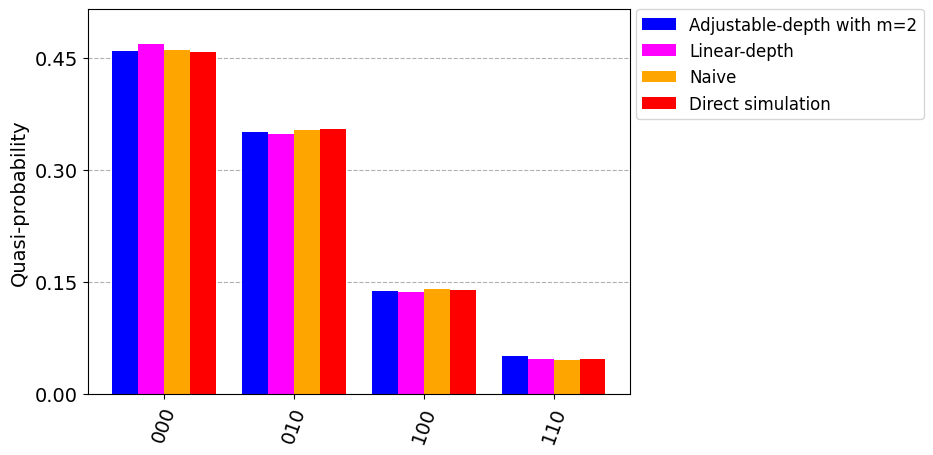} \ \ \
\includegraphics[width=8.5cm]{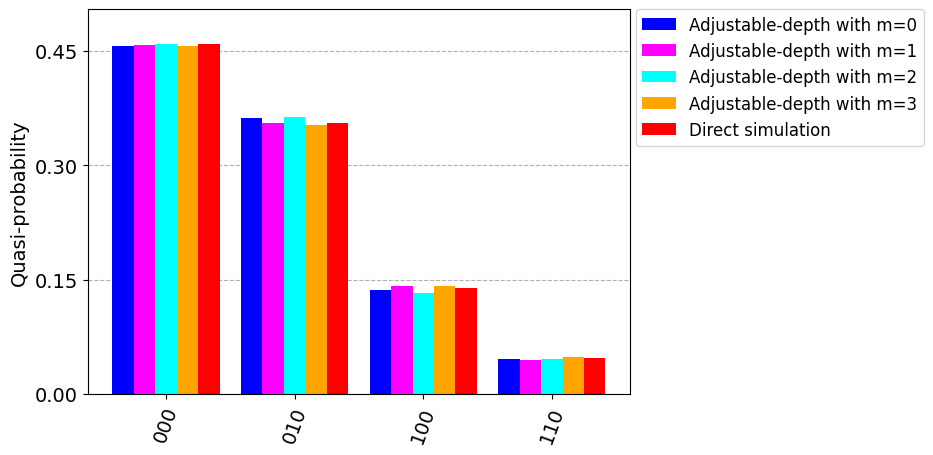}
\caption{\label{fig:histogram} Final probability distribution obtained by running $10000$ times a given circuit with $n=3$ position qubits up to $100$ time steps. In the left figure, we show the histograms for the naive circuit and the linear-depth circuit of Ref.\ \cite{NZDPplus2022}, for our adjustable-depth circuit with $m=2$, and for a direct simulation of the walk (i.e., without mapping it onto a quantum circuit), which we have called ``Matrix simulation'' in Ref.\ \cite{NZDPplus2022} instead of ``Direct simulation''; the angles used for the different coin operators are given in Table \ref{table:table}. In the right figure, we show the histograms for different choices of the adjustable parameter $m=0,1,2,3$.}
\end{figure*}

In Fig.\ \ref{fig:histogram}, we show the probability distribution obtained after $100$ time steps of running different circuits, with coin operators parametrized by
\begin{equation}
K(\alpha,\theta,\phi,\lambda) \defeq e^{i\alpha}
\begin{bmatrix}
\cos \frac{\theta}{2} & - e^{i\lambda} \sin \frac{\theta}{2} \\
e^{i\phi} \sin \frac{\theta}{2} &  e^{i(\phi+\lambda)} \cos \frac{\theta}{2} 
\end{bmatrix} \, ,
\label{eq:coin_op_new}
\end{equation}
with angles taken at random for each coin operators $C_i$, in the intervals
\begin{equation}
\alpha,\theta \in [0, \pi[ \ \text{and} \ \phi, \lambda \in [-\pi, \pi[ \, .
\end{equation}
The values obtained are given in Table \ref{table:table}.

\begin{figure*}
\includegraphics[width=6cm]{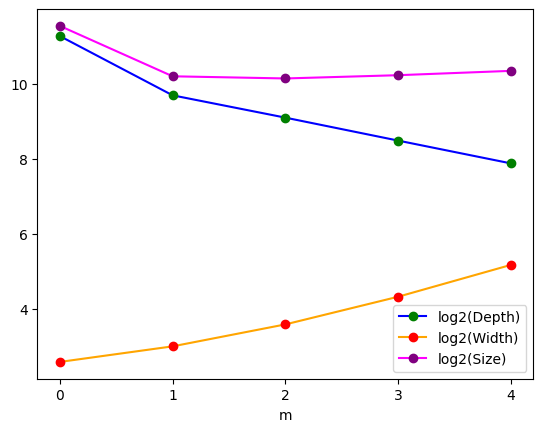} \ \ \ \
\includegraphics[width=6cm]{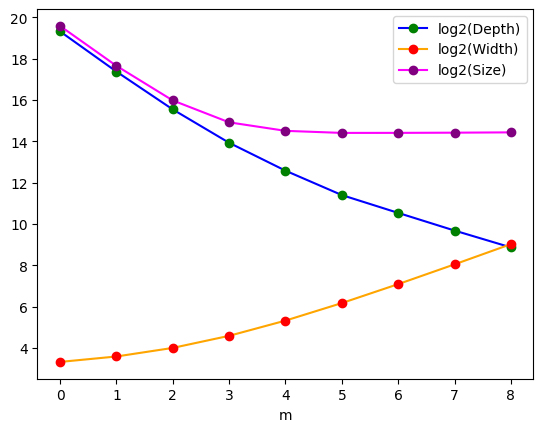}
\caption{\label{fig:depth_width} Depth, width, and size of the adjustable-depth quantum circuit, after compilation, for $n=4$ (left plots) and $n=8$ (right plots), as a function of the adjustable parameter $m$.}
\end{figure*}

In Fig.\ \ref{fig:depth_width}, we show, as a function of $m$, the depth, width and size of our adjustable-depth quantum circuit, after compilation by QASM.
The size is the number of one- and two-qubit gates involved in the circuit.
The compilation has been done with the following set of universal gates:
\begin{subequations}
\begin{align}
R_X(\theta) &\defeq
\begin{bmatrix}
\cos \frac{\theta}{2} & -i \sin \frac{\theta}{2} \\
 -i \sin \frac{\theta}{2} & \cos \frac{\theta}{2}
\end{bmatrix} \label{eq:RX}\\
R_Y(\theta) &\defeq
\begin{bmatrix}
\cos \frac{\theta}{2} & - \sin \frac{\theta}{2} \\
 \sin \frac{\theta}{2} & \cos \frac{\theta}{2}
\end{bmatrix} \\
R_Z(\lambda) &\defeq
\begin{bmatrix}
e^{-i \frac{\lambda}{2}} & 0 \\
0 & e^{i \frac{\lambda}{2} }
\end{bmatrix} \\
P(\lambda)  &\defeq
\begin{bmatrix}
1 & 0 \\
0 & e^{i {\lambda}} 
\end{bmatrix} \\
CNOT& \defeq
\begin{bmatrix}
1 & 0 & 0 & 0 \\
0 & 1 & 0 & 0 \\
0 & 0 & 0 & 1 \\
0 & 0 & 1 & 0 
\end{bmatrix} \, .
\end{align}
\end{subequations}
As $m$ increases, the depth decreases but the width increases, which is expected (the decrease of the depth being the purpose of increasing $m$, and the increase of the width being the consequence of needing to add ancillary wires in order to decrease the depth). Moreover, it is interesting to note the following non-trivial behavior: the number of gates decreases as $m$ increases, which may come from the fact that, as $m$ increases, the number of multi-Toffoli gates decreases (indeed, these multi-Toffoli gates have a high cost in terms of number of one- and two-qubit gates needed to implement them).

\begin{figure}
\begin{tabular}{|c|c|c|c|c|}
\hline
& $\alpha$ & $\theta$ & $\phi$ & $\lambda$ \\ \hline
$C_0$ & 2.59157236 & 0.07621657 &  2.38136754 & -2.79936369 \\ \hline
$C_1$ & 0.99031147& 0.27887516 & -2.67278043 &  1.84368898 \\ \hline
$C_2$ & 0.46354727 & 3.01620087 &  1.5039485 &  2.87444318 \\ \hline
$C_3$ & 1.79814059 & 2.5016676 & -0.50529796 & -1.65451052 \\ \hline
$C_4$ & 2.24708944 & 0.90254105 & -1.48779474 & -0.43149381 \\ \hline
$C_5$ & 1.99400865 & 0.7635951 & -2.15664402 & -0.07448489 \\ \hline
$C_6$ & 2.69049595 & 1.05540144 & -3.12609191 &  0.22005922 \\ \hline
$C_7$ & 2.77026061 & 1.2182012 & -0.29889881 & -0.72954279 \\ \hline
\end{tabular}
\caption{\label{table:table} Values of the angles of the coin operators, chosen for our implementation of the different circuits in Fig.\ \ref{fig:histogram}. The parametrization is given in Eq.\ \eqref{eq:coin_op_new}.}
\end{figure}

\section{Conclusions and discussion}
\label{sec:conclusions}

In this paper, we have provided a family of quantum circuits that implement a DQW on the line, with the following characteristics.
This family is parametrized by $m \in \mathbb{N}$.
For $m=0$, the circuit coincides with the naive circuit of Ref.\ \cite{NZDPplus2022}, which means that all the coin operators at each site of the line are implemented sequentially.
For $m=n$, the number of qubits used to encode the position of the walker, the circuit coincides with the linear-depth circuit of Ref.\ \cite{NZDPplus2022}, which means that all the coin operators are implemented in parallel.
A circuit with a given arbitrary $m$ means that the circuit contains $2^{n-m}$ packs, each of them containing $2^m$ coin operators implemented in parallel.
A higher (lower) $m$ means that more (less) coin operators are implemented in parallel, so that one can choose $m$ as best suited for the experimental platform, knowing that a higher $m$, and hence a smaller depth, requires more ancillary qubits.

It would be interesting to characterize the specific properties of DQWs, such as strict locality, at the level of their quantum-circuit translation \cite{PC20}.
It surely would be very interesting to extend the results of Ref.\ \cite{NZDPplus2022} and of this paper to quantum cellular automata (QCAs), which are multiparticle generalizations of DQWs \cite{Arrighi2019, Farrelly2020}.
Given that there are already some QCAs which simulate quantum electrodynamics in $1+1$, $1+2$ and $1+3$ dimensions \cite{ABF20, SADM2022, EDMMplus22}, this would mean having a quantum circuit which simulates some quantum field theory while implementing the strict locality of the transport.
If more properties or symmetries of the continuum model are preserved by the QCA that simulates it, such as Lorentz symmetry \cite{AFF14a, BDAP16, Debbasch2019a, Debbasch2019b} or gauge invariance \cite{ABF20, SADM2022, EDMMplus22}, such a quantum-circuit translation program would provide quantum circuits for quantum field theories, which respect many of the symmetries of the continuum model, which is not only a guarantee of numerical  accurateness, but also provides an alternate, lattice definition of these theories, endowed with all the symmetries required by physical principles, thus creating a new, lattice paradigm for such physical theories, phrased in terms of quantum circuits and thus directly implementable on most-used quantum hardware.
Finally, a question which is interesting is the following. Imagine that a certain algorithm is conceived with quantum walks, say, DQWs, and we want to run it with a quantum circuit. If we use the known algorithms that translate a DQW into a quantum circuit, do we obtain an algorithm that is as efficient as the original one made with a DQW, or do we have to modify it to reach the original efficiency?

\begin{center}

{\bfseries STATEMENT OF ABSENCE OF CONFLICT OF INTEREST}

\end{center}

On behalf of all authors, the corresponding authors state that there is no conflict of interest.

\begin{center}

{\centering \bfseries{DATA AVAILABILITY}}

\end{center}

Data will be made available upon reasonable request.



%

\appendix

\section{Explicit definition of $Q_{1,i}^{(n,m)}$}
\label{app:Q1}

\begin{figure*}
  \includegraphics[width=7cm]{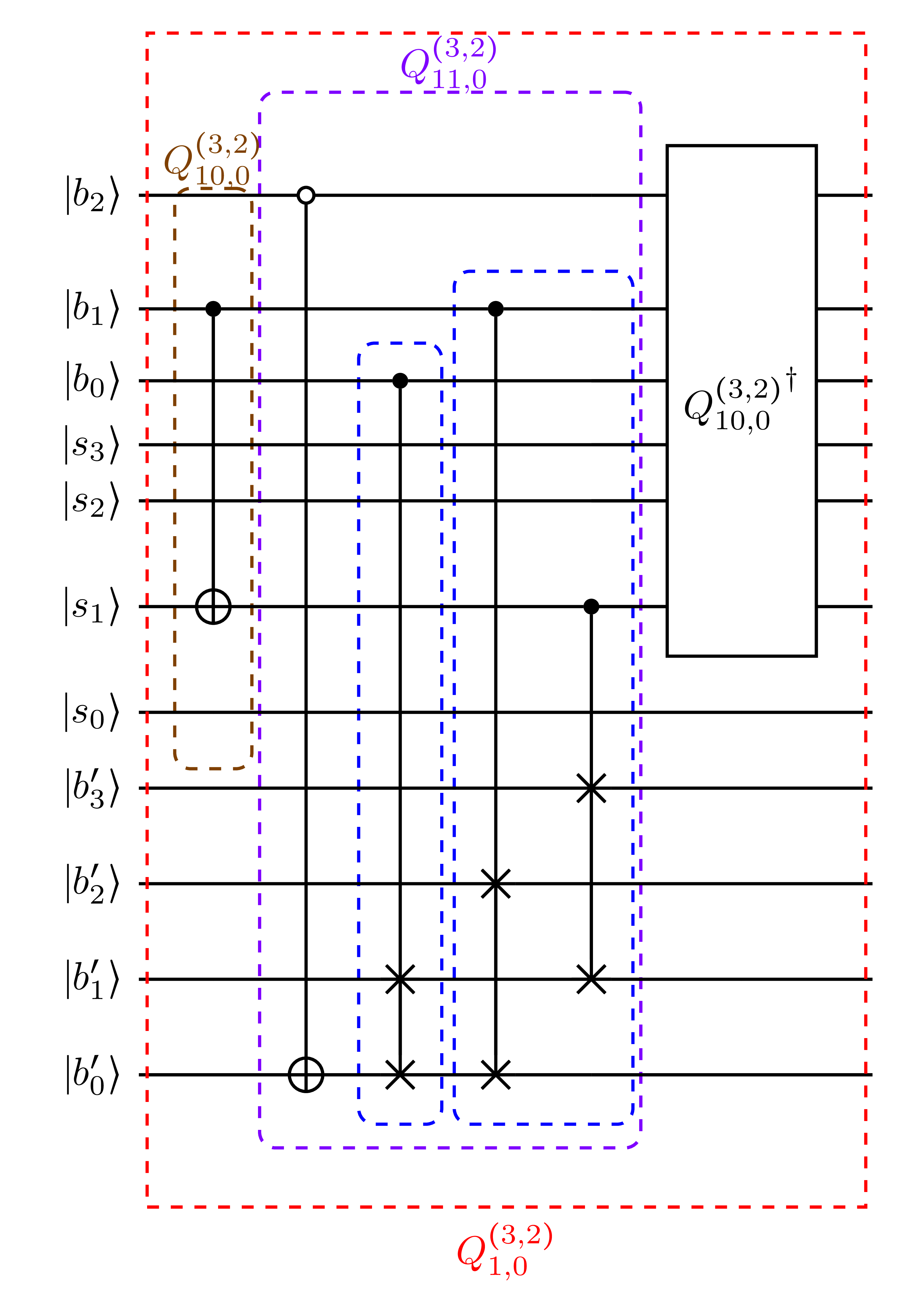} \ \ \ \
  \includegraphics[width=7cm]{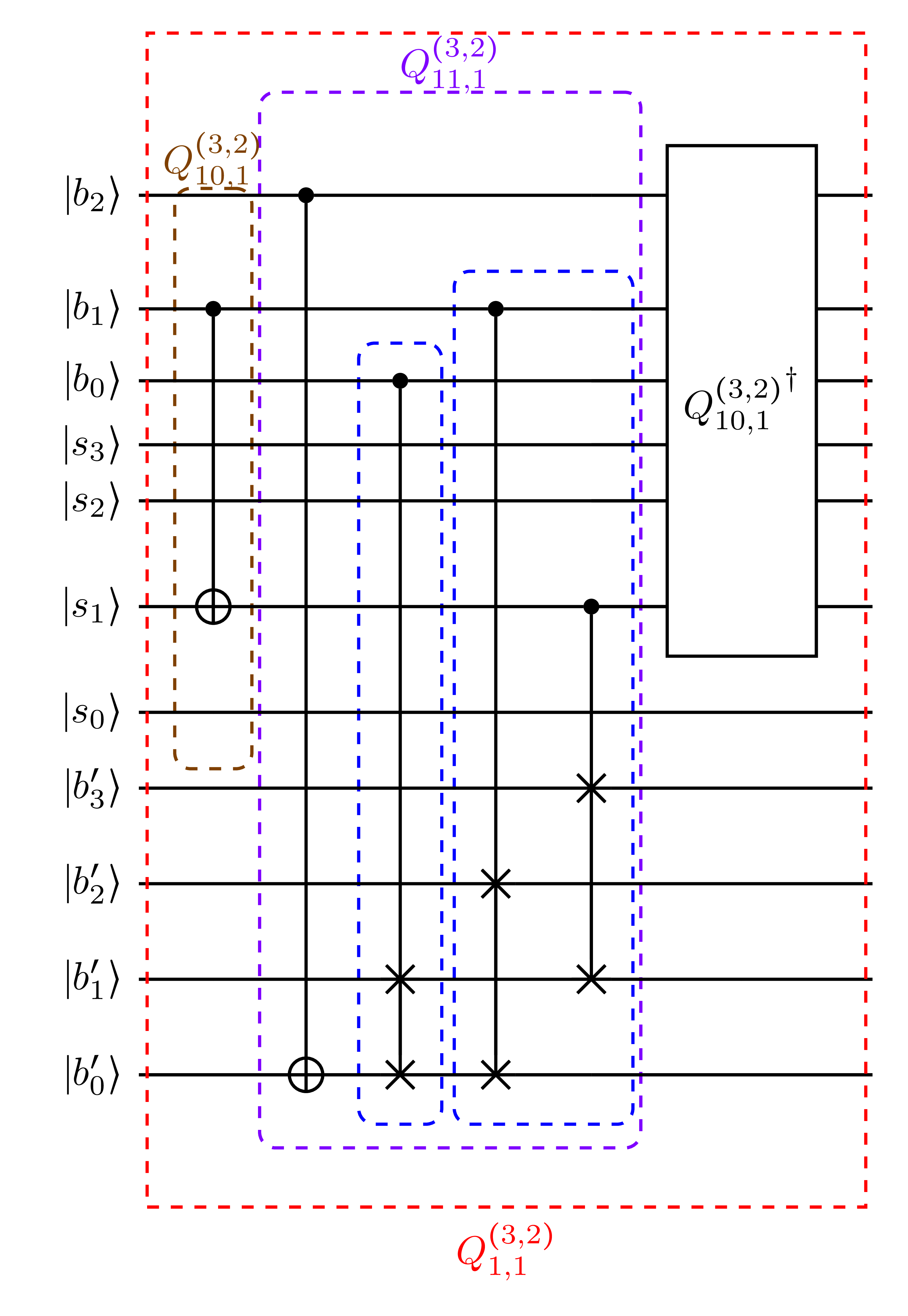}
\caption{Quantum circuits implementing $Q_{1,i}^{(n=3,m=2)}$ for $i=0,1$. \label{fig:q1}}
\end{figure*}

As mentioned in Sec.\ \ref{subsec:new_ingredient}, the only difference between $Q_{1,i}^{(n,m)}$ and $Q_1^{(n)}$ of Ref.\ \cite{NZDPplus2022} is, apart from the number of ancillary qubits, the fact that the first NOT gate of $Q_{11}^{(n)}$ is replaced by a generalized $(n-m)$-Toffoli gate in $Q_{11,i}^{(n,m)}$ (that we are going to define). 
To describe $Q_{1,i}^{(n,m)}$ explicitly, we thus follow exactly the construction of $Q_1^{(n)}$ in Ref.\ \cite[Appendix C3]{NZDPplus2022}.
Thus, the explicit definition of operator $Q_{1,i}^{(n,m)}$ can be written
\begin{equation}
\label{eq:q1}
Q_{1,i}^{(n,m)} \defeq {Q_{10}^{(n,m)}}^{\dag} Q_{11,i}^{(n,m)}Q_{10}^{(n,m)} \, ,
\end{equation}
where $Q_{10}^{(n,m)}$ makes the copies of the values of the position qubits on the ancillary coins, in order to be able to perform the controlled SWAPs in parallel via $Q_{11,i}^{(n,m)}$, and then one undoes the copies via ${Q_{10}^{(n,m)}}^\dag$.
The amount of copies and SWAPs is no longer quantified by $n$ as in Ref.\ \cite{NZDPplus2022}, but by the parameter $m$.
On Fig.\ \ref{fig:q1}, we have depicted the quantum circuits implementing $Q_{1,i}^{(n,m)}$ for $n=3$ and $m=2$, so that $i=0,1$.
Let us now write explicitly the copies operation, $Q_{10}^{(n,m)}$, and the controlled-SWAPs operation, $Q_{11,i}^{(n,m)}$.

\subsection{The copies: $Q_{10}^{(n,m)}$}

We have
\begin{equation}
    \label{eq:q10nm}
    Q_{10}^{(n,m)} \defeq \sideset{}{^L}\prod_{j=0}^{m-2}q_{10}^{(n,m)}(j) \, ,
\end{equation}
where $\forall j \geq 0$,
\begin{equation}
    \label{eq:18}
    q_{10}^{(n,m)}(j) \defeq \sideset{}{^L}\prod_{k=j+1}^{m-1}J_k^{(j)}\, ,
\end{equation}
with
{\small
\begin{subequations}
\begin{align}
    \label{eq:19}
    J_k^{(0)} &\defeq K_{b_k,s_{l_k}}(X) \\
    J_k^{(j\geq 1)} &\defeq \left(\sideset{}{^L}\prod_{l'=0}^{2^j-2}K_{s_{l_k}+l',s_{l_k}+l'+2^j}(X)\right)K_{b_k,s_{_{l_k}}+\sum_{u=1}^j2^{u-1}}(X) \, ,
\end{align}
\end{subequations}}
where we have defined
\begin{equation}
\begin{split}
    l_0 &\coloneqq 1 \\
    l_k &\coloneqq 2^{k-1}-1+l_{k-1},\forall k \geq 1 \, ,
\end{split}
\end{equation}
and where we remind the reader that $K_{a,b}(C)$ corresponds to applying the one-qubit gate $C$ on qubit $\ket b$ while controlling it on qubit $\ket a$ (we apply $C$ only if $a=1$).

\subsection{The controlled SWAPs: $Q_{11,i}^{(n,m)}$}

The operator $Q_{11,i}^{(n,m)}$ is composed of the generalized $(n-m)$-Toffoli gate, that we denote by $T_i^{(n,m)}$, and of the controlled-SWAP operations.
We can write it as
\begin{equation}
\label{eq:q11_decomposition}
    Q_{11,i}^{(n,m)} \defeq \Bigg(\underbrace{\sideset{}{^L}\prod_{j=0}^{m-1}B_j}_{\text{controlled SWAPs}}\Bigg) T_{i}^{(n,m)}\, ,
\end{equation}
where we are going to define $T_{i}^{(n,m)}$ and the $B_j$'s below.

\subsubsection{The generalized multi-Toffoli gate}

What we call generalized multi-Toffoli gate is a multi-Toffoli gate for which the controls can be positive (i.e., on $1$) or negative (i.e., on $0$).
The question here is how to express a negative control in terms of a positive control.

A first, naive idea is to express a negative control by the sequence of a NOT gate, then, a positive control, and finally another NOT gate.
But, have in mind that we do not apply $Q_{11,i}^{(n,m)}$ alone, we apply the Hermitian conjugate ${Q_{11,i}^{(n,m)}}^\dag$ later on.
Moreover, the $n-m$ last position qubits, on which the multi-Toffoli gate is controlled, are used by no other operation within $U_i^{(n,m)}$, neither the copies $Q_{10,i}^{(n,m)}$, nor the controlled SWAPs of $Q_{11,i}^{(n,m)}$, nor $Q_2^{(n,m)}$.
So, what we can do is simply applying a NOT gate before applying a positive control in order to obtain a negative control, and then we just wait until the application of ${Q_{11,i}^{(n,m)}}^\dag$ to undo these operations on the $n-m$ last position qubits.
One thus has to flip the qubit, i.e., to (i) apply a NOT gate on  $\ket{b_j}$, with $j=m,\dots, n-1$, whenever the bit $h_{j-m}$ of $i_2$ is equal to 0, and then to (ii) apply a positive control, which delivers in total a negative control on $\ket{b_j}$.  

Thus, we replace $T_i^{(n,m)}$ by
\begin{equation}
\label{eq:x}
\resizebox{.99\hsize}{!}{$
   \tilde{T}_{i}^{(n,m)} \defeq \underbrace{{K}_{\alpha,b'_0}^{\text{multi}}(X)}_{\text{$(n-m)$-Toffoli}} \left(\underbrace{\left(\bigotimes_{k=m}^{n-1}g_{i,k-m}^{b_{k}}\right)}_{\text{bit flips}}\otimes I_{2^m}\otimes I_{2^{(2^m)}}\otimes I_{2^{(2^m)}}\right) \, ,$}
\end{equation}
which we call almost generalized multi-Toffoli gate.
In this equation, the function $g^{b_k}_{i,j}$ indicates when to place the NOT gates on the control position qubit $\ket{b_k}$:
\begin{equation}\label{eq:g}
    g_{i,j}^{b_k}\defeq
    \begin{cases}
    X \text{ if } \lfloor \frac{i}{2^j} \! \! \! \mod 2 \rfloor = 0 \\
    I_2 \text{ otherwise}
    \end{cases} \, .
\end{equation}
Moreover, ${K}_{\alpha,b'_0}^{\text{multi}}(C)$ is the multiply controlled operation that applies gate $C$ on qubit $\ket{b'_0}$ whenever all qubits are $1$ in the set of $n-m$ control qubits
\begin{equation}
    \alpha \defeq \bigcup_{k=m}^{n-1} \{\ket{b_{k}}\} \, .
\end{equation}

In Appendix \ref{app:flip}, we present an alternative method for reducing the amount of NOT gates used in the generalized multi-Toffoli gate.

\subsubsection{The controlled SWAPs}

We can write
\begin{equation}\label{eq:bj}
    B_j \defeq \left(\bigotimes_{k=1}^{2^j-1}E_{b'_k,b'_k+2^j}^{s_{j+k-1}}\right)E_{b'_0,b'_{2^j}}^{b_j} \, ,
\end{equation}
where $E_{b,c}^{a}$ denotes the controlled-SWAP operation that swaps $\ket{b}$ and $\ket{c}$, controlling this by $\ket{a}$.

\section{Explicit definition of $Q_{2}^{(n,m)}$}
\label{app:Q2}

In this appendix, we give an explicit definition of $Q_{2}^{(n,m)}$, which is the same as that of $Q^{(n)}_2$ of Ref.\ \cite{NZDPplus2022} except for the number of ancillary wires.
In Fig.\ \ref{fig:q2}, we show the quantum circuits implementing $Q_{2}^{(n,m)}$ for $n=3$ and $m=0,1,2,3$.
The operation $Q_{2}^{(n,m)}$ can be written
\begin{equation}
\label{eq:q2}
    Q_{2}^{(n,m)} \defeq Q_{21}^{(n,m)}Q_{20}^{(n,m)} \, ,
\end{equation}
where $Q_{20}^{(n,m)}$ corresponds to the the starting series of CNOT operations, and $Q_{21}^{(n,m)}$ to the controlled-SWAP series followed by CNOTs.
The explicit definition of $Q_{20}^{(n,m)}$ reads
\begin{widetext}
\begin{equation}
\label{eq:q2A1}
    Q_{20}^{(n,m)} \defeq I_{2^n}\otimes I_{2^{(2^m)}}\otimes\left[ \sideset{}{^L}\prod_{j=0}^{m-2}\left(\bigotimes_{k=0}^{2^{m-j-1}-2}K_{b'_{2^m-1-2^{j+1}(\frac{1}{2}+k)},b'_{2^m-1-k2^{j+1}}}(X)\right)\right] \, ,
\end{equation}
while the explicit definition of $Q_{21}^{(n,m)}$ reads
\begin{equation}
\label{eq:q2A2}
    Q_{21}^{(n,m)} \defeq I_{2^n}\otimes\left[\sideset{}{^L}\prod_{j=0}^{m-1}C_{m,j}\left(\bigotimes_{k=0}^{2^j-1}E_{s_{k2^{m-j}},s_{2^{m-j}(\frac{1}{2}+k)}}^{b'_{(k+1)2^{m-j}-1}}\right)\right] \, ,
\end{equation}
where the CNOTs following the controlled SWAPs are given by
\begin{equation}
\label{eq:q2A3}
   C_{m,j} \defeq \left\{
   \begin{array}{ll}
\bigotimes_{l=0}^{2^{j+1}-2}K_{b'_{2^m-1-2^{m-j-1}(\frac{1}{2}+l)},b'_{2^m-1-l2^{m-j-1}}}(X) \ & \ \text{if} \ j < m-1 \\
I_{2^{(2^m)}} \otimes I_{2^{(2^m)}} \ & \ \text{if} \ j = m-1   
   \end{array} \right.
\end{equation}

\begin{figure*}
\hspace{-1cm}
\includegraphics[width=2cm]{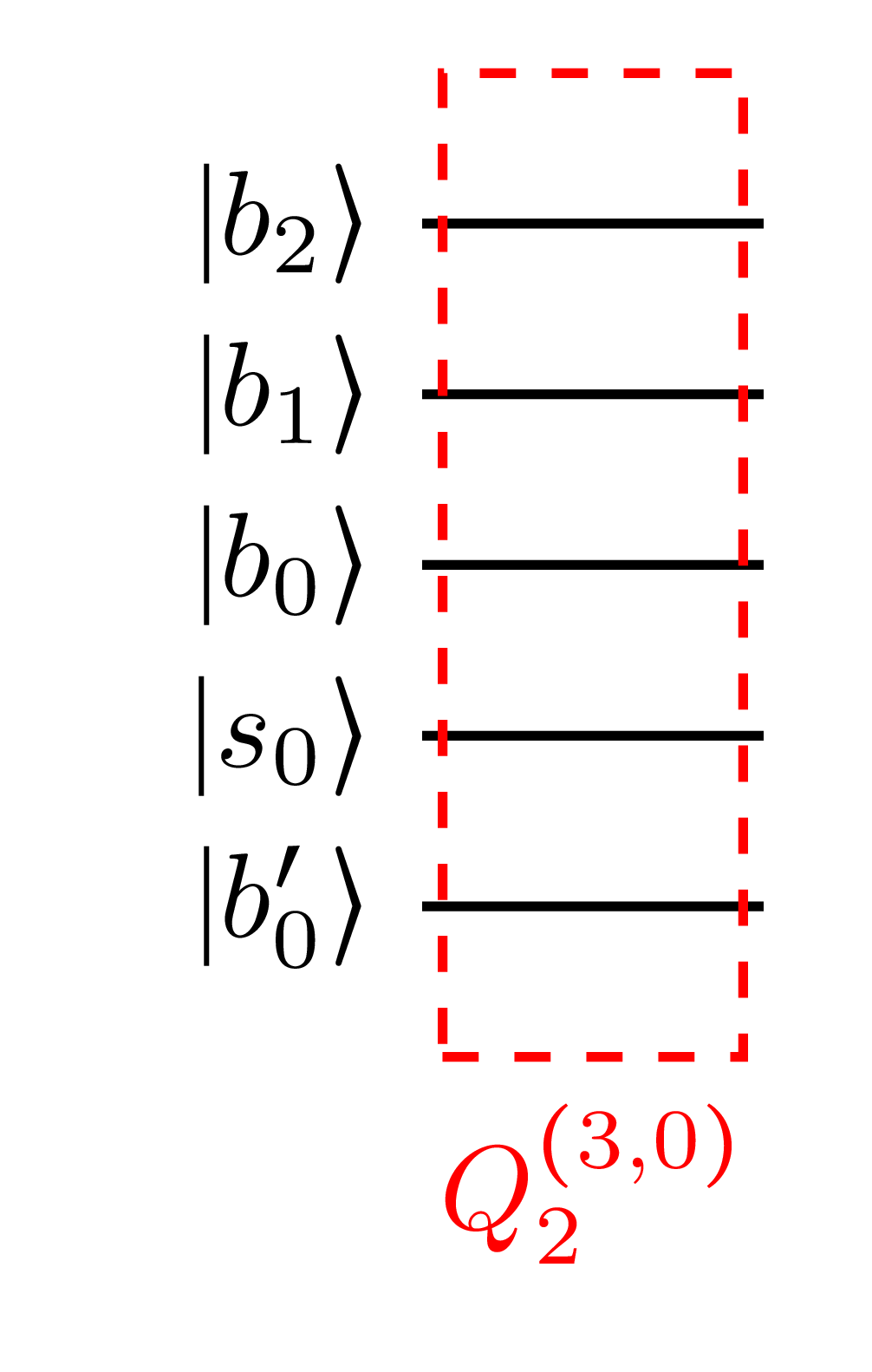}
\includegraphics[width=3cm]{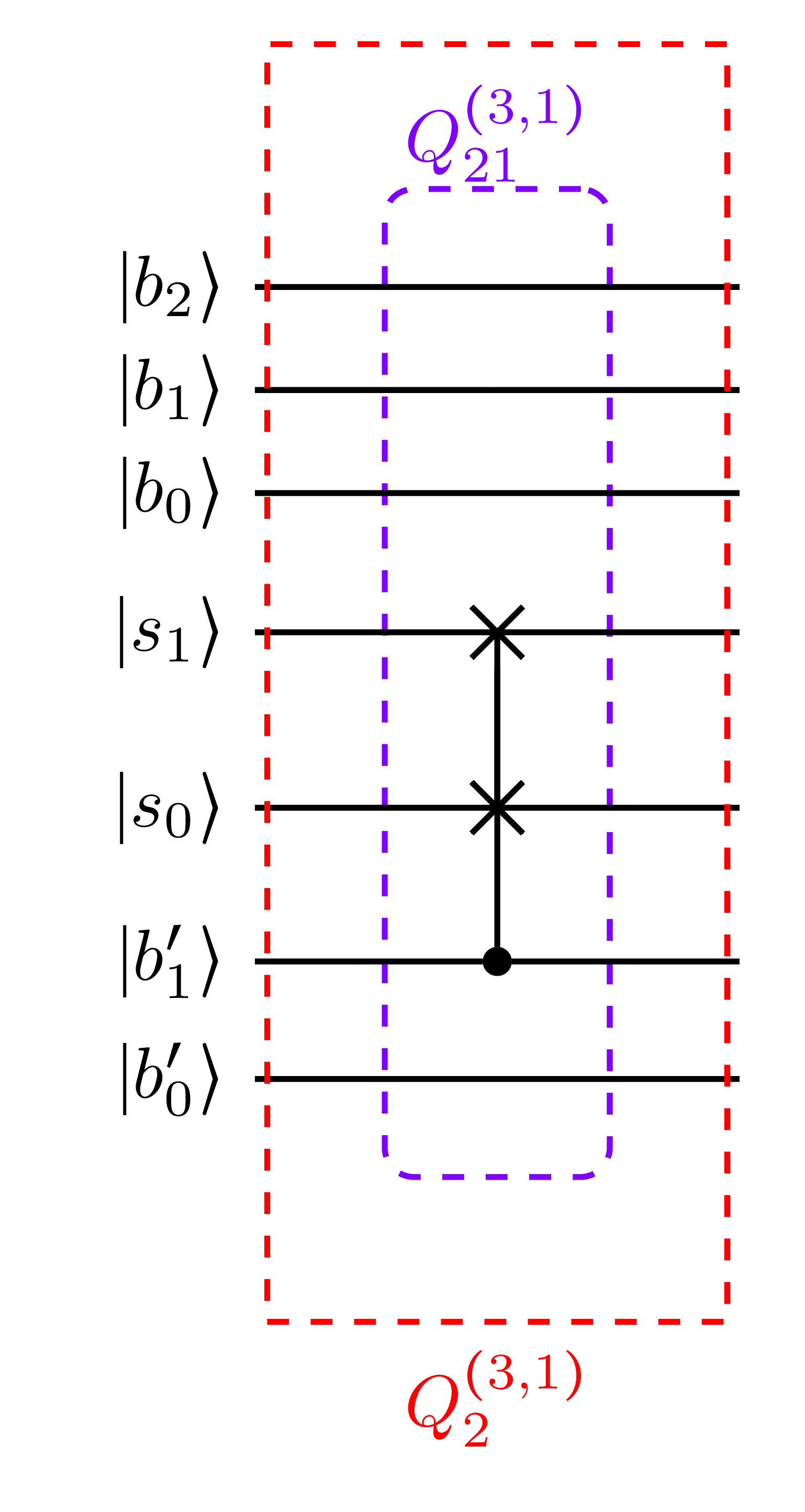}
\includegraphics[width=5cm]{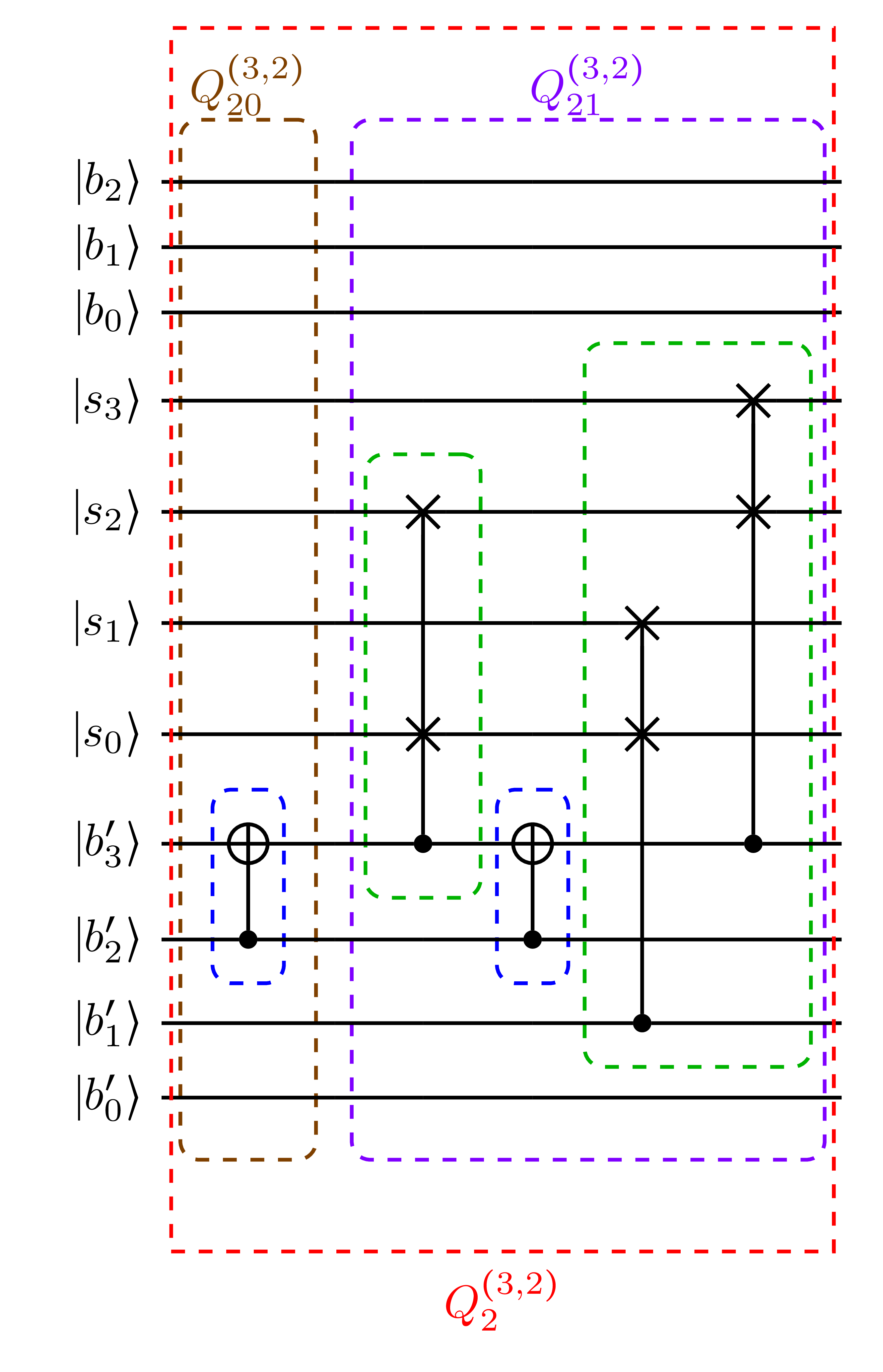}
\includegraphics[width=8cm]{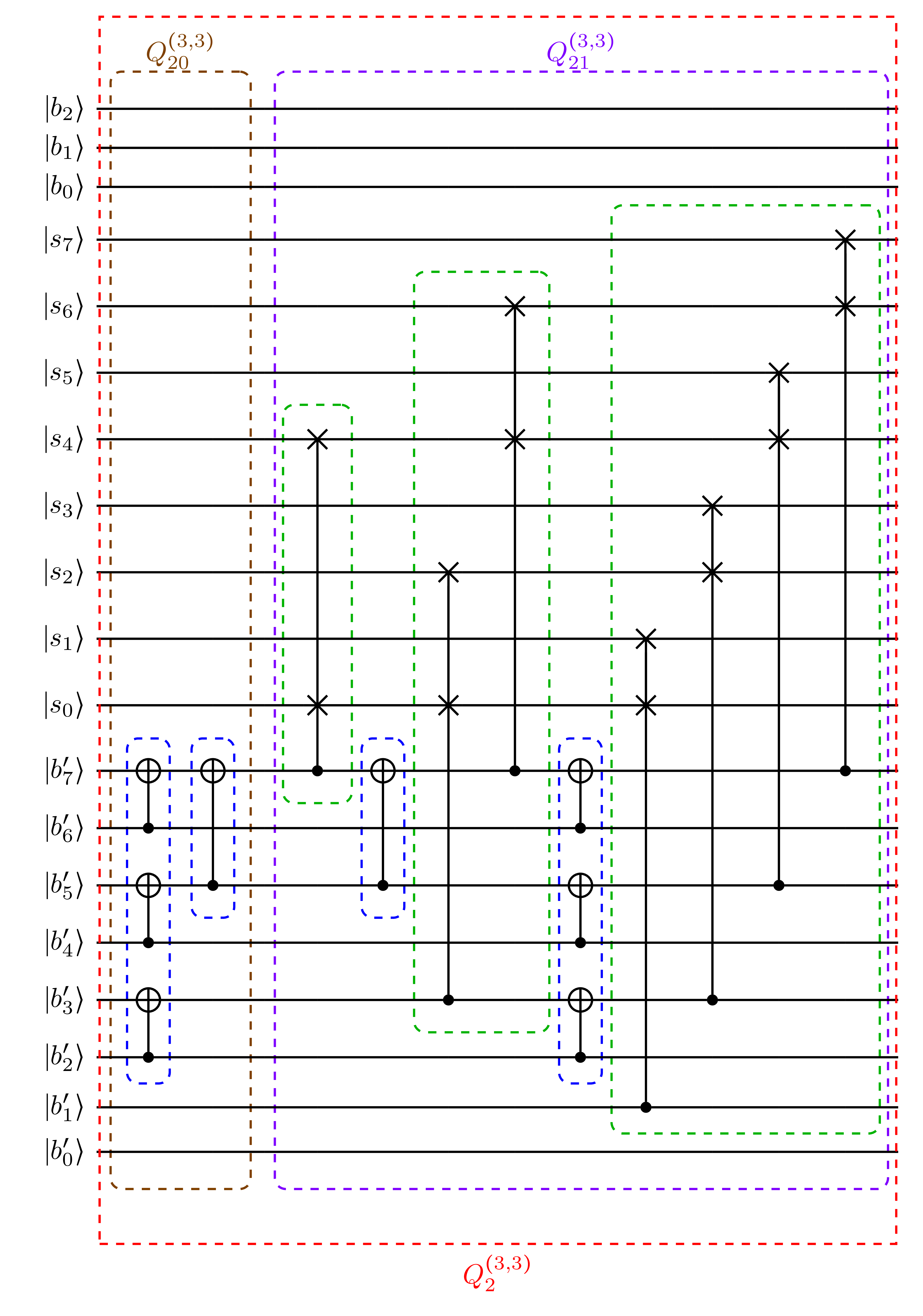}
\caption{Quantum circuits implementing $Q_{2}^{(n=3,m)}$, for $m=0,1,2,3$. \label{fig:q2}}
\end{figure*}

\section{Depth-calculation details}
\label{app:depth}

\begin{figure}[t]
\hspace{-0.0cm}
	\includegraphics[width=5cm]{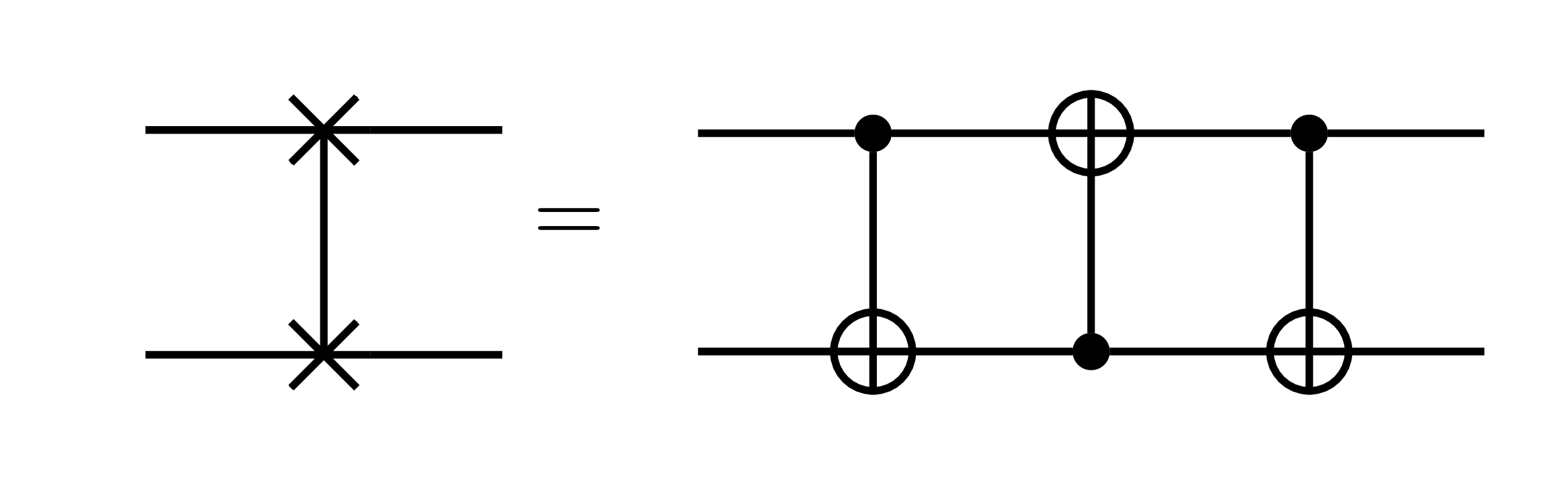}
	\caption{A way of expressing the SWAP operation, with 3 CNOT gates. \label{fig:swaps}}
\end{figure}

In this appendix, we prove the result for the depth of the circuit, given in Eq.\ \eqref{eq:depth}.

First, we recall that the depth of a SWAP operation counts for 3, as shown in Fig.\ \ref{fig:swaps}.

Now, the depth of $U^{(n,m)}$, defined in Eqs.\ \eqref{eq:packs} and \eqref{eq:packsQ}, is
\begin{subequations}
\begin{align}
\label{eq:u}
        d(U^{(n,m)}) &= \sum_{i=0}^{2^{n-m}-1}d(U_i^{(n,m)}) \\
        &= \sum_{i=0}^{2^{n-m}-1} d({Q_{1,i}^{(n,m)}}^{\dagger}) + d({Q_{2}^{(n,m)}}^{\dagger}) + d(Q_{0,i}^{(n,m)}) + d(Q_{2}^{(n,m)}) + d(Q_{1,i}^{(n,m)}) \, .
        \label{eq:sum}
\end{align}
\end{subequations}

Since we apply $2^m$ coin operators in parallel with $Q_{0,i}^{(n,m)}$, defined in Eq.\ \eqref{eq:q0} and illustrated in Fig.\ \ref{fig:Q0i}, we have that
\begin{equation}
\label{eq:depth_q0}
    d(Q_{0,i}^{(n,m)}) = 1 \, .
\end{equation}

The depth of the operator $Q_{2}^{(n,m)}$, defined in Eqs.\ \eqref{eq:q2}, \eqref{eq:q2A1}, \eqref{eq:q2A2} and \eqref{eq:q2A3}, and illustrated in Fig.\ \ref{fig:q2}, is
\begin{equation}
\begin{split}
    d(Q_{2}^{(n,m)}) &= 3\underbrace{m}_{\text{SWAPs}} + \underbrace{2(m-1)}_{\text{CNOTs}} \\
    &= 5m-2 \, .
\end{split}
\end{equation}
If $m=0$, we get a negative depth for $d(Q_{2}^{(n,m)})$, and the minimum value of the depth must be 0. To solve this problem we perform the following modification,
\begin{equation}
\label{eq:dq2}
d(Q_{2}^{(n,m)}) = 5m - 2(1-\delta_{m,0}) \, .
\end{equation}
Moreover, we have that 
\begin{equation}
\label{eq:equalityQ2}
d(Q_{2}^{(n,m)})  = d({Q_{2}^{(n,m)}}^\dagger)  \, .
\end{equation}

Lastly, the depth of the operator $Q_{1,i}^{(n,m)}$, defined in Eqs.\ \eqref{eq:q1}, \eqref{eq:q10nm} and \eqref{eq:q11_decomposition}, and illustrated in Fig.\ \ref{fig:q1}, is,
\begin{equation}\label{eq:dq1}
    d(Q_{1,i}^{(n,m)}) = d({Q_{10,i}^{(n,m)}}^{\dagger}) + d(Q_{11,i}^{(n,m)}) + d(Q_{10,i}^{(n,m)}) \, ,
\end{equation}
where
\begin{equation}
\label{eq:equality}
    d(Q_{10,i}^{(n,m)}) = d(Q_{10,i}^{(n,m)\dagger}) \, ,
\end{equation}
and
\begin{equation}
    d(Q_{10,i}^{(n,m)}) = m-1 \, .
\end{equation}
If $m=0$ we obtain a negative depth for  $d(Q_{10,i}^{(n,m)})$; we therefore perform the following modification,
\begin{equation}
\label{eq:dq10}
    d(Q_{10,i}^{(n,m)}) = m-1+\delta_{m,0} \, .
\end{equation}
As shown in Eq.\ \eqref{eq:q11_decomposition}, the operator $Q_{11,i}^{(n,m)}$ can be separated into two operations: the $(n-m)$-Toffoli gate on $\ket{b'_0}$, and the series of controlled SWAPs. Let us first treat the $(n-m)$-Toffoli gate. As mentionned in Sec.\ \ref{subsec:new_ingredient}, one has to flip some of the position qubits before applying the $(n-m)$-Toffoli gate; the only pack $i$ for which no flip is needed is when $i=2^{n-m}-1$, i.e., the last pack; therefore, we get a contribution $1-\delta_{i,2^{n-m}-1} + \varepsilon_d(n-m)$ to $d(Q_{11,i}^{(n,m)})$, where we recall that $\varepsilon_d(n-m)$ denotes the depth of the $(n-m)$-Toffoli gate. Let us now treat the controlled SWAPs. The depth of the non-parallelized controlled SWAPs is $3\sum_{k=0}^{m-1}2^k=3(2^m-1)$; however, as one parallelizes this step in the circuit, the depth becomes only $3m$. Therefore, the depth of $Q_{11,i}^{(n,m)}$ finally reads
\begin{equation}
\label{eq:dq11}
    d(Q_{11,i}^{(n,m)}) = 1-\delta_{i,2^{n-m}-1} + \varepsilon_d(n-m) + 3m \, .
\end{equation}
Inserting Eqs.\ \eqref{eq:equality}, \eqref{eq:dq10} and \eqref{eq:dq11} into Eq.\ \eqref{eq:dq1}, one gets,
\begin{subequations}
\begin{align}
         d(Q_{1,i}^{(n,m)}) &= d(Q_{10,i}^{(n,m)\dagger}) + d(Q_{11,i}^{(n,m)}) + d(Q_{10,i}^{(n,m)}) \\
         &= 2d(Q_{10,i}^{(n,m)}) + d(Q_{11,i}^{(n,m)}) \\
         &= 2(m-1+\delta_{m,0}) +  1-\delta_{i,2^{n-m}-1} + \varepsilon_d(n-m) + 3m \\
         &= 5m + \varepsilon_d(n-m) + 2\delta_{m,0} - \delta_{i,2^{n-m}-1} - 1 \, .
         \label{eq:dq1i}
\end{align}
\end{subequations}
Moreover, we have that 
\begin{equation}
\label{eq:equalityQ1i}
d(Q_{1,i}^{(n,m)})  = d({Q_{1,i}^{(n,m)}}^\dagger)  \, .
\end{equation}

Inserting now Eqs.\ \eqref{eq:equalityQ2} and \eqref{eq:equalityQ1i}, and then \eqref{eq:depth_q0},  \eqref{eq:dq2} and \eqref{eq:dq1i}, into the expression of $d(U^{(n,m)}_i)$ given by Eq.\ \eqref{eq:sum}, we obtain
\begin{subequations}
\begin{align}
        d(U_i^{(n,m)}) &= d(Q_{1,i}^{(n,m)\dagger}) + d(Q_{2}^{(n,m)\dagger}) + d(Q_{0,i}^{(n,m)}) + d(Q_{2}^{(n,m)}) + d(Q_{1,i}^{(n,m)}) \\
    &= 2(d(Q_{2}^{(n,m)}) + d(Q_{1,i}^{(n,m)})) + d(Q_{0,i}^{(n,m)}) \\
    &= 2(5m - 2(1-\delta_{m,0}) + 5m + \varepsilon_d(n-m) + 2\delta_{m,0} - \delta_{i,2^{n-m}-1} - 1) + 1 \\
    &= 20m + 2\varepsilon_d(n-m) - 2\delta_{i,2^{n-m}-1} + 8\delta_{m,0} - 5 \, .
    \label{eq:finalll}
\end{align}
\end{subequations}

The only term of $d(U_i^{(n,m)})$ which depends on $i$ is $-2\delta_{i,2^{n-m}-1}$, which is equal to -2 when $i=2^{n-m}-1$ and 0 for the rest; thus, inserting  Eq.\ \eqref{eq:finalll} into Eq.\ \eqref{eq:u}, we get
\begin{subequations}
\begin{align}
        d(U^{(n,m)}) &= \sum_{i=0}^{2^{n-m}-1}d(U_i^{(n,m)}) \\
        &= \sum_{i=0}^{2^{n-m}-1}20m + 2\varepsilon_d(n-m) - 2\delta_{i,2^{n-m}-1} + 8\delta_{m,0} - 5 \\
        &= (2^{n-m}-1)(20m + 2\varepsilon_d(n-m)+8\delta_{m,0}-5) + 20m + 2\varepsilon_d(n-m) + 8\delta_{m,0} - 7 \\
        &= 2^{n-m}(20m+2\varepsilon_d(n-m) + 8\delta_{m,0}-5)-2 \, ,
\end{align}
\end{subequations}
which is the result announced in Eq.\ \eqref{eq:depth}.
\end{widetext}

\section{Optimizing the number of NOT gates used in $Q_{1,i}^{(n,m)}$}
\label{app:flip}

The NOT gates applied in order to realize the almost generalized multi-Toffoli gate $\tilde{T}_i^{(n,m)}$ (see Eq.\ \eqref{eq:x}), i.e., applied with the function $g^{b_k}_{i,j}$ before the standard multi-Toffoli gate, are applied again when applying the conjugate transposed ${Q_{1,i}^{(n,m)}}^{\dag}$, and part of these NOT gates of $(\tilde{T}_i^{(n,m)})^{\dag}$ cancel out with the NOT gates applied in order to realize the next almost generalized multi-Toffoli gate  $\tilde{T}_{i+1}^{(n,m)}$.
Therefore, it makes sense to devise a function that only applies the NOT gates remaining after the cancelling out.
More precisely, this function replaces  $g^{b_k}_{i,j}$, and is applied only to implement $\tilde{T}_i^{(n,m)}$, i.e., only before the standard multi-Toffoli gate of $Q_{11,i}^{(n,m)}$, and not after applying the same standard multi-Toffoli gate of ${Q_{11,i}^{(n,m)}}^{\dag}$.
This is indeed possible because the operations which are
in between $(\tilde{T}_i^{(n,m)})^{\dag}$ and $\tilde{T}_{i+1}^{(n,m)}$, namely, the conjugate transposed of the copies ${Q_{10}^{(n,m)}}^{\dag}$ and the copies ${Q_{10}^{(n,m)}}$ (which by the way simplify each other apart from the last stage), do not involve the wires on which $(\tilde{T}_i^{(n,m)})^{\dag}$ controls.

So, in each $U_i^{(n,m)}$, we do the following modifications: (i) instead of applying $\tilde{T}_i^{(n,m)}$ in $Q_{1,i}^{(n,m)}$ (see Eq.\ \eqref{eq:x}), we apply the same operation but replacing $g_{i,j}^{b_k}$ given in Eq.\ \eqref{eq:g} by
\begin{equation}
h_{i,j}^{b_k} \defeq
    \begin{cases}
    X \ \ \text{if $ i \text{\hspace{-2mm}}\mod 2^j = 0$} \\
    I_2 \ \text{ otherwise}
    \end{cases}  ,
\end{equation}
which amounts to replacing $Q_{1,i}^{(n,m)}$ by an operation that we call $P_{1,i}^{(n,m)}$; (ii)
moreover, instead of applying $(\tilde{T}_i^{(n,m)})^{\dag}$, we simply apply the standard $(n-m)$-Toffoli gate ${K}_{\alpha,b'_0}^{\text{multi}}(X)$, which amounts to replacing ${Q_{1,i}^{(n,m)}}^{\dag}$ by an operation that we call $\bar{P}_{1,i}^{(n,m)}$.
In total, we have replaced $U_i^{(n,m)}$ by
\begin{equation}
\label{eq:Uprime}
(U_i^{(n,m)})' \defeq  \bar{P}_{1,i}^{(n,m)} {Q_{2}^{(n,m)}}^\dagger Q_{0,i}^{(n,m)}Q_{2}^{(n,m)}{P}_{1,i}^{(n,m)} \, .
\end{equation}

\begin{figure*}
\includegraphics[width=18cm]{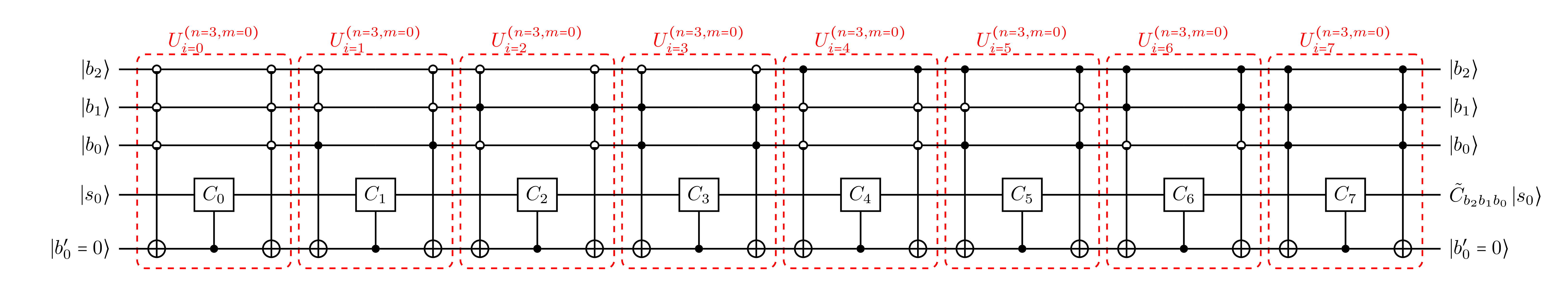}
\includegraphics[width=18cm]{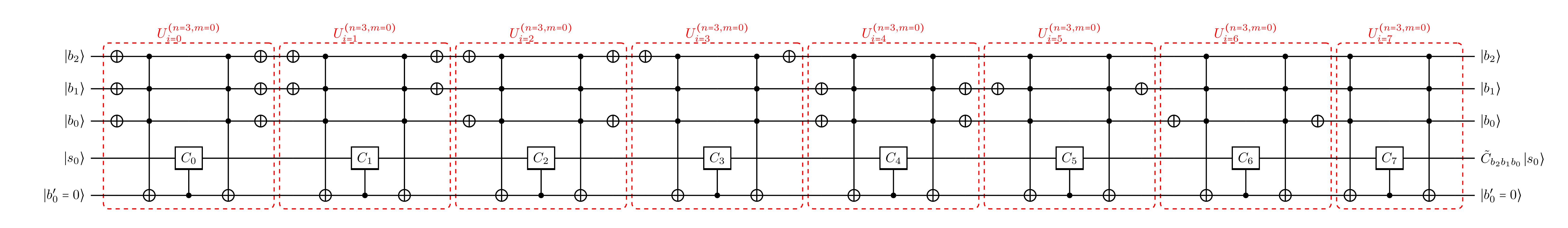}
\includegraphics[width=18cm]{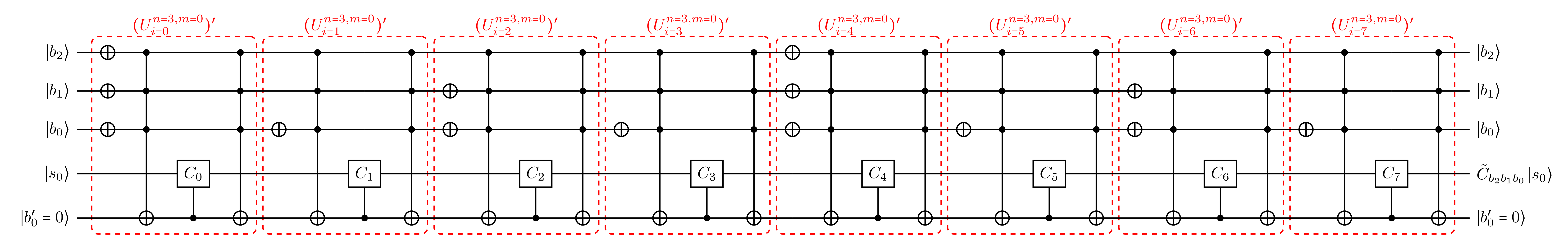}
\caption{ In the top figure, we show $U^{(n=3,m=0)}$. In the middle figure, we replace the generalized multi-Toffoli gates by the $\tilde{T}_i^{(n,m)}$'s, which enables to see that part of the NOT gates of  the $(\tilde{T}_i^{(n,m)})^\dag$'s cancel out with those of  the $\tilde{T}_{i+1}^{(n,m)}$'s. In the bottom figure, we show the optimized circuit $(U^{(n,m)})' $, obtained by taking the product of the $(U_i^{(n,m)})'$'s, defined in Eq.\ \eqref{eq:Uprime}. \label{fig:optimization}}
\end{figure*}

Let us notice that $h_{i,j}^{b_k}$ implements the NOT gates as in the naive circuit of Ref.\ \cite{NZDPplus2022}.
In Fig.\ \ref{fig:optimization}, we show how part of the NOT gates of the almost generalized multi-Toffoli gates $(\tilde{T}_i^{(n,m)})^\dag$ and $\tilde{T}_{i+1}^{(n,m)}$ cancel between each other, and which NOT gates remain, that we encode via $h_{i,j}^{b_k}$.

\section{Pseudo-code}
\label{app:pseudo_code}

The pseudo-code used to code, with Qiskit, the adjustable-depth quantum circuit, is given below:
\begin{algorithm}[H]
\caption{$Q_{1,i}^{(n,m)}$ and $P_{1,i}^{(n,m)}$}
\begin{algorithmic}
\State $Q_{10}^{(n,m)}$:
\For{$j \in [0\dots m-2]$}
\For{$k \in [j+1\dots m-1]$}
\If{$j=0$}
\State $not(s_{l_k}).control(b_k)$
\Else
\State $not(s_{l_k+\sum_{u=1}^j2^{u-1}}).control(b_k)$
\For{$p \in [0\dots 2^j-2]$}
\State $not(s_{l_k+p+2^j}).control(s_{l_k+p})$
\EndFor
\EndIf
\EndFor
\EndFor \\
\State $Q_{11,i}^{(n,m)}$ and $P_{11,i}^{(n,m)}$:
\State $\alpha\gets \{b_k | k \in [m\dots n-1]\}$
\For{$k \in [m\dots n-1]$}
\If{$P_{11,i}^{(n,m)}$}
\If{$i \mod 2^{k-m}=0$}
\State $not(b_k)$
\EndIf
\Else
\If{$\lfloor i/2^{k-m} \mod 2 \rfloor = 0$}
\State $not(b_k)$
\EndIf
\EndIf
\EndFor
\If{$m\neq n$}
\State $not(b'_0).control(\alpha)$
\Else
\State $not(b'_0)$
\EndIf
\For{$j\in [0\dots m-1]$}
\State $swap(b'_0,b'_{2^j}).control(b_j)$
\For{$k\in [1\dots 2^j-1]$}
\State $swap(b'_k,b'_{2^{k+2^j}}).control(s_{j+k-1})$
\EndFor
\EndFor \\
\State $\bar{P}_{11,i}^{(n,m)}$:
\State $\alpha\gets \{b_k | k \in [m\dots n-1]\}$
\If{$m\neq n$}
\State $not(b'_0).control(\alpha)$
\Else
\State $not(b'_0)$
\EndIf
\For{$j\in [0\dots m-1]$}
\State $swap(b'_0,b'_{2^j}).control(b_j)$
\For{$k\in [1\dots 2^j-1]$}
\State $swap(b'_k,b'_{k+2^j}).control(s_{j+k-1})$
\EndFor
\EndFor
\end{algorithmic}
\end{algorithm}

\begin{algorithm}[H]
\caption{$Q_{2}^{(n,m)}$}
\begin{algorithmic}
\For{$j \in [0\dots m-2]$}
\For{$k \in [0\dots 2^{m-j-1}-2]$}
\State $not(b'_{2^m-1-k2^{j+1}}).control(b'_{2^m-1-2^{j+1}(1/2+k)})$
\EndFor
\EndFor
\For{$j \in [0\dots m-1]$}
\For{$k \in [0\dots 2^{j}-1]$}
\State $swap(s_{k2^{m-j}}, s_{(1/2+k)2^{m-j}}).control(b'_{(k+1)2^{m-j}-1})$
\EndFor
\If{$j \neq m-1$}
\For{$l \in [0\dots 2^{j+1}-2]$}
\State $not(b'_{2^m-1-l2^{m-j-1}}).control(b'_{2^m-1-2^{m-j-1}(1/2+l)})$
\EndFor
\EndIf
\EndFor
\end{algorithmic}
\end{algorithm}

\begin{algorithm}[H]
\caption{$Q_{0,i}^{(n,m)}$}
\begin{algorithmic}
\For{$k \in [0\dots 2^m-1]$}
\State $C_{i2^m+k}(s_k,b'_k)$
\EndFor
\end{algorithmic}
\end{algorithm}


\end{document}